\documentclass[9.25pt,twocolumn,final]{IEEEtran}

\usepackage[colorlinks,linkcolor=black, urlcolor=black, citecolor=black]{hyperref}
\usepackage{amsmath}
\usepackage{amssymb}
\usepackage{amsmath,bm}
\usepackage{enumerate}
\usepackage{subfig}
\usepackage{graphicx}
\usepackage{cite}
\usepackage{caption}
\usepackage{algorithm}
\usepackage{algorithmic}
\usepackage{booktabs}
\usepackage{array}
\usepackage{multirow}
\usepackage{mathrsfs}
\usepackage{txfonts}
\usepackage{amsfonts,amssymb}
\usepackage{makecell}
\usepackage{pifont}

\usepackage{epstopdf}
\usepackage{epsfig}
\usepackage{fancyhdr}

\begin{document}
\title{An Overview of Cross-media Retrieval: Concepts, Methodologies, Benchmarks and Challenges}
\author{Yuxin Peng,
        Xin Huang,
        and Yunzhen Zhao
\thanks{This work was supported by National Natural Science Foundation of China under Grants 61371128 and 61532005.

The authors are with the Institute of
Computer Science and Technology, Peking University, Beijing 100871,
China. Corresponding author: Yuxin Peng (e-mail: pengyuxin@pku.edu.cn).}}
\maketitle

\thispagestyle{fancy}
\fancyhead{} 
\lhead{} 
\cfoot{Copyright~\copyright~2017 IEEE. Personal use of this material is permitted. However, permission to use this material for any other purposes must be obtained from the IEEE by sending an email to pubs-permissions@ieee.org.} 
\rfoot{}

\begin{abstract}
Multimedia retrieval plays an indispensable role in big data utilization. Past efforts mainly focused on single-media retrieval. However, the requirements of users are highly flexible, such as retrieving the relevant audio clips with one query of image. So challenges stemming from the ``media gap", which means that representations of different media types are inconsistent, have attracted increasing attention. Cross-media retrieval is designed for the scenarios where the queries and retrieval results are of different media types. 
As a relatively new research topic, its concepts, methodologies and benchmarks are still not clear in the literatures. 
To address these issues, we review more than 100 references, give an overview including the concepts, methodologies, major challenges and open issues, as well as build up the benchmarks including datasets and experimental results. 
Researchers can directly adopt the benchmarks to promptly evaluate their proposed methods. This will help them to focus on algorithm design, rather than the time-consuming compared methods and results.
It is noted that we have constructed a new dataset \emph{XMedia}, which is the first publicly available dataset with up to five media types (text, image, video, audio and 3D model).
We believe this overview will attract more researchers to focus on cross-media retrieval and be helpful to them.

%To address these issues, this paper reviews more than 100 references, gives an overview including the concepts, methodologies as well as major challenges and open issues, and also builds up the benchmarks including datasets and experiment results, aiming to facilitate the research of cross-media retrieval. 
%Researchers can directly adopt the datasets and results to promptly evaluate their proposed methods. This will help them to focus on algorithm design, rather than the time-consuming compared methods and results.
% It should be noted that we have constructed a new dataset XMedia, which is the first publicly available dataset with up to 5 media types (text, image, video, audio and 3D model) to the best of our knowledge. 
% We believe that this overview will attract more researchers to focus on cross-media retrieval and be helpful to them.

\end{abstract}
\IEEEpeerreviewmaketitle
\section{Introduction}
\label{section:introduction}
\IEEEPARstart{W}{ith} the rapid growth of multimedia data such as text, image, video, audio and 3D model, cross-media retrieval is becoming increasingly attractive, through which users can get the results with various media types by submitting one query of any media type. For instance, on a visit to the  Gate Bridge, users can submit a photo of it, and retrieve relevant results including text descriptions, images, videos, audio clips and 3D models. 

%one query of any media type such as an image, and retrieve relevant results of all media types. % as shown in Figure \ref{fig:cmrExample}.
%, such as text descriptions, images, videos, and audio clips, which is shown  in Figure \ref{fig:cmrExample}.
%For example, user can submit an image, and get not only the relevant images, but also texts, videos, audio clips and even 3D models as results.

The research of multimedia retrieval has lasted for several decades \cite{LewTOMCCAP06MIRReview}. However, past efforts generally focused on single-media retrieval, where the queries and retrieval results belong to the same media type.
Beyond the case of single-media retrieval, some methods have been proposed to deal with more than one media type. Such methods aim to combine multiple media types together in a retrieval process as \cite{clinchant2011semantic, LiuCIVR2010coherent}, but the queries and retrieval results must share the same media combination. For example, users can retrieve image/text pairs by an image/text pair. Although these methods involve multiple media types, they are not designed for performing retrieval across different media types, and cross-media similarities cannot be directly measured, such as the similarity between an image and an audio clip.
Nowadays, as digital media content is generated and found everywhere,  requirements of users are highly flexible such as retrieving the relevant audio clips with one query of image. Such retrieval paradigm is called cross-media retrieval, which has been drawing extensive interests. It is more useful and flexible than single-media retrieval because users can retrieve whatever they want by submitting whatever they have \cite{YangMM09LRGA}. %As a new retrieval paradigm, cross-media retrieval has been drawing extensive interests \cite{LiuIJMIS2010CMRSurvey}. 

\begin{figure*}[t]
\begin{center}
   \includegraphics[width=0.93\linewidth]{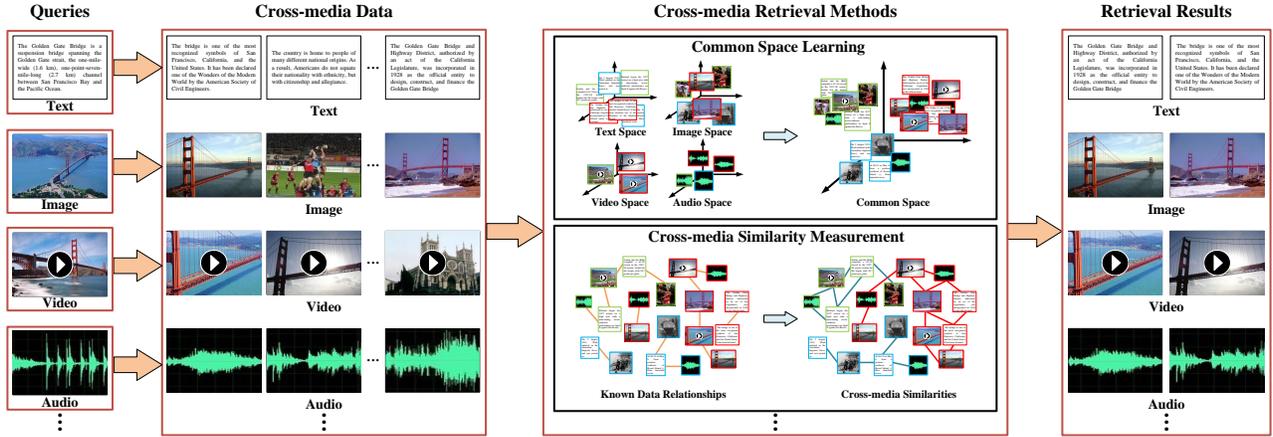}
\end{center}
   \caption{A brief illustration of cross-media retrieval, which shows two major kinds of methods, namely common space learning and cross-media similarity measurement methods. %For instance, users can submit an image, and get not only the relevant images, but also the relevant texts, videos, audios and even 3D models. %Cross-media retrieval provides results with multiple media types at the same time given.
   }
\label{fig:cmrExample}
\vspace{-3mm}
\end{figure*}

The key challenge of cross-media retrieval is the issue of ``media gap", which means that representations of different media types are inconsistent and lie in different feature spaces, so it is extremely challenging to measure similarities among them.
% (e.g., the similarity between an image of wolf and an audio clip of wolf howl).
There have been many methods proposed for addressing this issue by analyzing the rich correlations contained in cross-media data. For example, current mainstream methods are designed to learn an intermediate common space for features of different media types, and measure the similarities among them in one common space, which are called \emph{common space learning methods}. Meanwhile, \emph{cross-media similarity measurement methods} are proposed to directly compute the cross-media similarities by analyzing the known data relationships without obtaining an explicit common space. A brief illustration of cross-media retrieval is shown in Figure \ref{fig:cmrExample}.  Most of the existing methods are designed for retrieval of only two media types (mainly image and text), but cross-media retrieval emphasizes the diversity of media types. Hence, there still remains a problem of incorporating other media types into the unified framework, such as video, audio and 3D model.
%which can be further applied to other media types.

As our research on cross-media retrieval has lasted for several years \cite{ZhaiMMM2012HSNN, ZhaiICASSP2012CCP, ZhaiAAAI2013JGRHML, MaICIP2013CBCA,  DBLP:journals/mms/ZhaiPX13, ZhaiTCSVT2014JRL, PengHypergraph2015,  DBLP:conf/ijcai/PengHQ16}, we find some key issues on concepts, methodologies and benchmarks are still not clear in the literatures. %, which has caused the confusions and limitations. 
To address these problems, we review more than 100 references and aim to:

\label{section:intr}

\begin{itemize}
\item Summarize existing works and methodologies to present an overview, which will facilitate the research of cross-media retrieval.
\item Build up the benchmarks, including datasets and experimental results. This will help researchers to focus on algorithm design, rather than the time-consuming compared methods and results, since they can directly adopt the benchmarks to promptly evaluate their proposed methods.
% and facilitate the development of cross-media retrieval
%Researchers can directly adopt the datasets and results to promptly evaluate their proposed methods. This will help them to focus on algorithm design, rather than the time-consuming compared methods and results.
\item Provide a new dataset \emph{XMedia} for comprehensive evaluations of cross-media retrieval. It is the first publicly available dataset consisting of up to five media types (text, image, video, audio and 3D model).
\item Present the main challenges and open issues, which are important and meaningful for the further research directions of cross-media retrieval.
\end{itemize}

The rest of this paper is organized as follows: Section \ref{section:problem} presents the definition of cross-media retrieval. Sections \ref{section:representation}, \ref{section:similarity} and \ref{section:other} introduce the representative works of common space learning, cross-media similarity measurement and other methods, which are shown in Table \ref{tab:Literature}. Section \ref{section:dataset} summarizes the widely-used datasets for cross-media retrieval, and Section \ref{section:experiment} presents the  experimental results on these datasets. Section \ref{section:challenges} presents the open issues and challenges, and finally Section \ref{section:conclusion} concludes this paper.

%\begin{figure}[t]
%\begin{center}
%   \includegraphics[width=\linewidth]{Fig/literaturemap.eps}
%\end{center}
%   \caption{A literature map of the representative cross-media retrieval works for better understanding. Tag \textbf{U} is for un-supervised method, \textbf{S} is for semi-supervised method and \textbf{F} is for fully-supervised method. Some methods involve relevance feedback as an important part, and cannot be easily categorized according to supervision settings, so they are given the tag \textbf{R}. Note that the classification is according to the main part of the literatures. %For example, in the common space learning section, we will only focus on the main process of common space learning, although some literatures may contain other processes.
  % For example, in section II some methods learn the common space under unsupervised settings, and then train classifiers under supervised settings for further retrieval. We will treat them as unsupervised settings because the key part we are discussing about.
  %}
%\label{fig:literMap}
%\end{figure}

\section{Definition of Cross-media Retrieval}
\label{section:problem}

%Cross-media retrieval methods depending on \emph{co-existence information} exclusively are regarded as un-supervised methods. On the contrary, if the training process 
%methods are regarded as supervised if the model is trained with labeled data \emph{semantic category labels} are taken for training as supervision information, no matter whether \emph{co-existence information} is used or not. Among supervised methods, fully-supervised methods only take labeled data for training, while semi-supervised methods can simultaneously exploit labeled and unlabeled data.

%Now we present the formulation of definition for cross-media retrieval. 
For clarity, we take two media types $X$ and $Y$ as examples to give the formulation of definition for cross-media retrieval. The training data is denoted as $\mathcal{D}_{tr}=\{{X}_{tr} , {Y}_{tr}\}$, in which ${X}_{tr}=\{{{\bm x}_p }\}_{p=1}^{n_{tr}}$, where  ${n_{tr}}$ denotes the number of media instances for training, and ${\bm x}_p$ denotes the $p$-th media instance. Similarly, we denote ${Y}_{tr}=\{{{\bm y}_p }\}_{p=1}^{n_{tr}}$. 
There exist co-existence relationships between ${\bm x}_p$ and ${\bm y}_p$, which mean that instances of different media types exist together to describe relevant semantics. %So there are totally $n_{tr}$ pairs of media types ${X}$ and ${Y}$. 
Semantic category labels for training data can be provided and denoted as $\{c^X_p\}_{p=1}^{n_{tr}}$ and $\{c^Y_p\}_{p=1}^{n_{tr}}$, which indicate the semantic categories that media instances belong to.
The testing data is denoted as  $\mathcal{D}_{te}=\{{X}_{te} , {Y}_{te}\}$, in which ${X}_{te}=\{{{\bm x}_q }\}_{q=1}^{n_{te}}$, and ${Y}_{te}=\{{{\bm y}_q }\}_{q=1}^{n_{te}}$.
The goal is to compute cross-media similarities $sim({\bm x}_a, {\bm y}_b)$, and retrieve relevant instances of different media types in testing data for one query of any media type.
%In this section, we present the definition of cross-media retrieval. As mentioned in Section \ref{section:introduction}, the aim is to bridge the ``media gap" for performing retrieval across different media types. %For obtaining the cross-media retrieval model, training information is needed. 
%There are mainly two kinds of information to guide the model training in existing works: \emph{co-existence information} means that instances of different media types exist together to describe relevant semantics, such as the relevance between an image and its text description; \emph{semantic category labels} indicate the semantic categories that media instances belong to.
Unsupervised methods take the setting that all training data is unlabeled, semi-supervised methods take the setting that only a subset of the training data is labeled, while fully supervised methods take the setting that all of the training data is labeled.
\renewcommand{\arraystretch}{1.1}
\begin{table}[t]
  \caption{Representative works of cross-media retrieval. \textbf{U} is for unsupervised methods, \textbf{S} is for semi-supervised methods, \textbf{F} is for fully supervised methods, and \textbf{R} is for methods that involve relevance feedback and cannot be easily categorized by supervision settings.} %Note that the classification is according to the main part of the literatures.}
	\begin{center}
		\scalebox{0.8}{
		\begin{tabular}{|c|c|c|} 
			\hline
			\multicolumn{2}{|c|}{Categories}& \multirow{1}{*}{Representative works} \\
			\hline
			\multirow{16}{*}{\begin{tabular}{c} Common \\ space\\ learning \end{tabular}} 
			&  {\begin{tabular}{c} Traditional statistical \\ correlation analysis \end{tabular}} 
			& {\begin{tabular}{c}  \cite{HotelingBiometrika36RelationBetweenTwoVariates}(U) \cite{DBLP:journals/neco/HardoonSS04}(U) \cite{Akaho01akernel}(U)  \cite{ClusterCCA}(F) \\ \cite{ICML2013DCCA}(U) \cite{DBLP:journals/ijcv/GongKIL14}(U,F) \cite{DBLP:conf/iccv/RanjanRJ15}(F)   \cite{RasiwasiaMM10SemanticCCA}(F)\\ \cite{costa2014role}(F) \cite{CVPR2012GMA}(F) \cite{LiMM03CFA}(U) \end{tabular}} \\
			\cline{2-3}
			
			&  {\begin{tabular}{c} DNN-based \end{tabular}} 
			& {\begin{tabular}{c} \cite{DBLP:conf/ijcai/PengHQ16}(F) \cite{ICML2013DCCA}(U) \cite{ngiam32011multimodal}(U) \cite{srivastava42012multimodal}(U) \\ \cite{DBLP:conf/cvpr/YanM15}(U) \cite{feng12014cross}(U) \cite{zhang2014start}(U) \cite{DBLP:conf/icml/WangALB15}(U) \\ \cite{DBLP:journals/tip/WuLSYZRZ16}(U) \cite{DBLP:journals/tcyb/WeiZLWLZY17}(F)
			\cite{DBLP:journals/tmm/HeXKWP16} (U)
			\cite{WangStackedAE2014}(U) \cite{DBLP:journals/corr/CastrejonAVPT16}(F) \end{tabular}} \\
			\cline{2-3}
						
			&  {\begin{tabular}{c} Cross-media graph \\ regularization \end{tabular}} 
			& {\begin{tabular}{c} \cite{ZhaiAAAI2013JGRHML}(F) \cite{ZhaiTCSVT2014JRL}(S) \cite{PengHypergraph2015}(S) \cite {DBLP:journals/tmm/WuYYTZZ14}[F] \\ \cite{PAMI15Graph}(S) \cite{DBLP:conf/sigir/LiangLCHW16}(U) \end{tabular}} \\
			\cline{2-3}
			
			&  {\begin{tabular}{c} Metric learning \end{tabular}} 
			& {\begin{tabular}{c} \cite{ZhaiAAAI2013JGRHML}(F) \cite{WuMetric2010}(F) \end{tabular}} \\
			\cline{2-3}
			
			&  {\begin{tabular}{c} Learning to rank \end{tabular}} 
			& {\begin{tabular}{c} \cite{GrangierPAMI08TBIR}(U) \cite{WuACMMM2013}(F) \cite{JiangDeep2015}(F) \cite{DBLP:journals/tip/WuJLTLZZ15}(F) \end{tabular}} \\
			\cline{2-3}

			&  {\begin{tabular}{c} Dictionary learning  \end{tabular}} 
			& {\begin{tabular}{c} \cite{NIPS2010DL}(U) \cite{ZhuDL2014}(S,F) \cite{zhuang2013supervised}(F) \\  \end{tabular}} \\
			\cline{2-3}
						
			&  {\begin{tabular}{c} Cross-media\\ hashing \end{tabular}} 
			& {\begin{tabular}{c} \cite{kumar2011learning}(U) \cite{zhang2011composite}(U) \cite{DBLP:journals/tmm/WuYYTZZ14}(F) \cite{zhen2012co}(U) \\
			 \cite{hu2014iterative}(F)  \cite{zhen2012probabilistic}(F) \cite{YuSIGIR2014DCDH}(F) \cite{DBLP:conf/sigir/LongCWY16}(U)	\\
			\cite{bronstein2010data}(F) \cite{rastegari2013predictable}(U) 	\cite{DingCollectiveHash2014}(U) \cite{ZhaiParametricHash2013}(F) \\
			\cite{zhuang2014cross}(F) \cite{DBLP:conf/ijcai/WangSS15}(U) 
			\cite{DBLP:conf/ijcai/WangGWH15}(U) \cite{DBLP:conf/ijcai/LiuJWH16}(F) \\
			\cite{DBLP:conf/ijcai/WangCO015}(U) \cite{ZhouLatentSemanticHash2014}(U) \end{tabular}} \\
			\cline{2-3}
			
			&  {\begin{tabular}{c} Others \end{tabular}} 
			& {\begin{tabular}{c} \cite{DBLP:conf/bmvc/VermaJ14}(U) \cite{ZhangSCP2014}(F) \cite{DBLP:journals/tmm/KangXLXP15}(F) \cite{HuaTina2014}(F) \\ \cite{DBLP:journals/tist/WeiZZWXFY16}(F) \cite{mao2013parallel}(S) \end{tabular}} \\
			\hline
			
			\multirow{2}{*}{\begin{tabular}{c} Cross-media \\ similarity\\ measurement \end{tabular}} 
			&  {\begin{tabular}{c} Graph-based \end{tabular}} 
			& {\begin{tabular}{c} \cite{YangMM09LRGA}(R) \cite{ZhaiICASSP2012CCP}(S) \cite{ZhuangTMM08SemanticCorrelation}(R) \cite{tong2005graph}(R) \\ \cite{DBLP:journals/pr/YangWXZC10}(R) \cite{YangIEEETPAMI2012}(R) \cite{zhuang2006approach}(R)  \cite{YangTMM08MMDManifolds}(R) \end{tabular}} \\
			\cline{2-3}
			
			&  {\begin{tabular}{c} Neighbor analysis \end{tabular}} 
			& {\begin{tabular}{c} \cite{clinchant2011semantic}(U) \cite{ZhaiMMM2012HSNN}(F) \\ \cite{MaICIP2013CBCA}(U,F) \end{tabular}} \\
			\hline
			
			\multirow{3}{*}{\begin{tabular}{c} Others \end{tabular}} 
						&  {\begin{tabular}{c} Relevance feedback \\analysis \end{tabular}} 
						& {\begin{tabular}{c} \cite{YangMM09LRGA}(R) \cite{ZhuangTMM08SemanticCorrelation}(R) \cite{YangIEEETPAMI2012}(R) \cite{YangTMM08MMDManifolds}(R)  \end{tabular}} \\
						\cline{2-3}
									&  {\begin{tabular}{c} Multimodal\\ topic model \end{tabular}} 
									& {\begin{tabular}{c} \cite{cLDA}(U) \cite{TrmmLDA}(U) \cite{JiaICCV2011MDRF}(U)  \cite{wang22014multi}(F) \end{tabular}} \\
									\cline{2-3}								
		%	&  \multicolumn{2}{c|}{\begin{tabular}{c} \cite{YangMM09LRGA}(R) \cite{ZhuangTMM08SemanticCorrelation}(R) \cite{YangIEEETPAMI2012}(R) \cite{YangTMM08MMDManifolds}(R) 
		%	\\ \cite{cLDA}(U) \cite{TrmmLDA}(U) \cite{JiaICCV2011MDRF}(U)  \cite{wang22014multi}(F) \end{tabular}} \\
			\hline
		\end{tabular} 
	}
	\end{center}
	  \label{tab:Literature}
	  \vspace{-3mm}
\end{table}

%If methods take the setting that training data is given with \emph{co-existence relationships} exclusively, they are regarded as un-supervised methods. Supervised methods take labeled data with \emph{semantic category labels} for training, no matter whether \emph{co-existence relationships} are used or not. 
%Among supervised methods, fully-supervised methods are designed for the setting that all training data is labeled, while for semi-supervised methods, only a subset of the training data is labeled.

Some works involve analyzing the correlation between different media types, mainly image and text%. For example, image annotation, image/video caption and cross-media topic detection. 
, but they are quite different from cross-media retrieval. For example, image annotation methods such as \cite{JeonSIGIR03ImageAnnotation} aim to obtain the probabilities that the tags are assigned to images, while in cross-media retrieval, text refers to sentences or paragraph descriptions rather than only tags. 
%Methods of image/video caption as \cite{DBLP:journals/corr/MaoXYWY14a, DBLP:conf/cvpr/VinyalsTBE15} mainly focus on generating descriptions for image/video, while cross-media retrieval aims to find the most relevant text for image/video in an existing dataset and vice versa. Another important difference between image/video caption and cross-media retrieval is that image/video caption only focuses on image/video and text, which is not easy to extend to other media types. 
Methods of image/video caption such as \cite{DBLP:journals/corr/MaoXYWY14a, DBLP:conf/cvpr/VinyalsTBE15} are mainly designed for generating the text descriptions of image/video, while cross-media retrieval aims to find the most relevant texts in the existing data for image/video in the existing data and vice versa. Another important difference between them is that image/video caption only focuses on image/video and text, which is not easy to be extended to other media types, while cross-media retrieval is the retrieval across all media types such as text, image, video, audio and 3D model. 
%As for cross-media topic detection \cite{ChuTD2014}, it aims to detect topics that exist in different media types simultaneously, and it is not a retrieval problem.
%In addition, transfer learning may be performed between image and text as \cite{DBLP:journals/tomccap/TangSLQW16}, but the aim is to exploit the data in source domain (like text domain) to benefit the task on data in target domain (like image domain). As indicated in \cite{DBLP:journals/tnn/ShaoZL15}, transfer learning mainly aims to extract the knowledge from one or more source tasks and applies the knowledge to a target task, and usually focus on the scenarios where there exist distinct source domain and target domain. While in cross-media retrieval, different media types are treated equally and all involved for retrieval, so there are no distinct source and target domains and tasks. 
In addition, there are some transfer learning works involving different media types such as \cite{DBLP:journals/tomccap/TangSLQW16}, but transfer learning and cross-media retrieval differ in two aspects:
(1) Transfer learning is a learning framework with a broad coverage of methods and applications, which allows the domains, tasks, and distributions used in training and testing to be different \cite{DBLP:journals/tkde/PanY10}. However, cross-media retrieval is a specific information retrieval task across different media types, and its characteristic challenge and focus are the issue of ``media gap". (2) ``Transfer learning aims to extract the knowledge from one or more source tasks and applies the knowledge to a target task" \cite{DBLP:journals/tkde/PanY10}, and there exist distinct source and target domains. But different media types are treated equally in cross-media retrieval, and there are usually no distinct source and target domains, or source and target tasks. % and the goal is to perform retrieval across all media types instead of only focusing on a certain target task. %Therefore, there are still essential differences between transfer learning and cross-media retrieval.

\section{Common Space Learning}
\label{section:representation}

Common space learning based methods are currently the mainstream in cross-media retrieval. They follow the idea that data sharing the same semantics has latent correlations, which makes it possible to construct a common space. %In this space, the relevant data will be close to each other, regardless of their media types.
Taking the Golden Gate Bridge as an example, all of the text descriptions, images, videos, audio clips and 3D models about it describe similar semantics. Consequently, they can be close to each other in a common high-level semantic space.
These methods aim to learn such a common space, and explicitly project data of different media types to this space for similarity measurement. 

We mainly introduce seven categories of existing methods as subsections \emph{A-G}. Among them, \emph{(A) traditional statistical correlation analysis methods} are the basic paradigm and foundation of common space learning methods, which mainly learn linear projection matrices for common space by optimizing the statistical values. Other categories are classified according to the characteristics on different aspects: 
\begin{itemize}

\item On \emph{\textbf{basic model}}, \emph{(B) DNN-based methods} take deep neural network as the basic model and aim to make use of its strong abstraction ability for cross-media correlation learning.
%\item \emph{(A) Traditional statistical correlation analysis methods} are the basic paradigms and foundations, which aim to optimize the certain statistical values, and mainly learn linear projection matrices. The other categories of methods achieve improvement in different aspects.
\item On \emph{\textbf{correlation modeling}},  \emph{(C) cross-media graph regularization methods} adopt the graph model to represent the complex cross-media correlations,  \emph{(D) metric learning methods} view the cross-media correlations as a set of similar/dissimilar constraints, and \emph{(E) learning to rank methods} focus on cross-media ranking information as their optimization objective.

\item On \emph{\textbf{property of common space}}, \emph{(F) dictionary learning methods} generate dictionaries and the learned common space is for sparse coefficient of cross-media data, and \emph{(G) cross-media hashing methods} aim to learn a common Hamming space to accelerate retrieval.
\end{itemize}

Because these categories are classified according to different aspects, there exist a few overlaps among these categories. For example, the work of \cite{ZhaiAAAI2013JGRHML} can be classified as both a metric learning and graph regularization method.
%For example, the work of \cite{DBLP:conf/ijcai/WangCO015} is both DNN-based and hashing method, and that of \cite{ZhaiAAAI2013JGRHML} is both metric learning and graph regularization method.

%There are some classical statistical correlation analysis methods for this scenario, such as CCA and CFA. By adding constraints on correlation and/or semantic similarity, common space learning can also be performed based on some existing subspace learning methods. They have some different names, such as semantic space and cross-media feature representation. However, they share the same idea of common space, and explicitly convert the media objects with different media types to representations of the same dimension. The representative works will be discussed below.

\subsection{Traditional Statistical Correlation Analysis Methods}
Traditional statistical correlation analysis methods are the basic paradigm and foundation of common space learning methods, which mainly learn linear projection matrices by optimizing the statistical values. Canonical correlation analysis (CCA) \cite{HotelingBiometrika36RelationBetweenTwoVariates} is one of the most representative works as introduced in \cite{DBLP:journals/neco/HardoonSS04}. The cross-media data is often organized as sets of paired data with different media types such as image/text pairs. CCA is a possible solution for such case, which learns a subspace that maximizes the pairwise correlations between two sets of heterogeneous data. As an early classical work, CCA is also used in some recent works such as \cite{DBLP:conf/bmvc/VermaJ14, DBLP:conf/cvpr/KleinLSW15}.
%According to the formula definition, we have ${X}_{Tr}=\{{{\bm x}_p }\}_{p=1}^{n_{Tr}}$ and ${Y}_{Tr}=\{{{\bm y}_p }\}_{p=1}^{n_{Tr}}$, and each ${\bm x}_p$ is corresponded with ${\bm y}_p$. For convenience, here we take $X$ for ${X}_{Tr}$, and $Y$ for ${Y}_{Tr}$. We let $d_x$ and $d_y$ denote the dimensional number of the feature vector for media types $X$ and $Y$, and the aim of CCA is to learn two projection matrices  $P_{x}\in\mathbb{R}^{d_{x} \times d_c}$ and $P_{y}\in\mathbb{R}^{d_{y} \times d_{c}}$
%in which $d^c$ denotes the dimension of the common space. The learning objective is to maximize the correlation as:
%\begin{align}
%    \rho = \max_{P_{x},P_{y}}corr(P_{x}^TX,P_{y}^TY)
%   = \max_{P_{x},P_{y}}\frac{{P_x}^TX{Y}^T{P_y}}{\left \|  {P_x}^TX\right \| \left \|  {P_y}^TY\right \|}
%\end{align}
%We observe the covariance matrix of (x,y) is:
%\begin{align}
%    C(x,y) = \left(
%    \begin{array}{cc}
%     C_{xx} & C_{xy}  \\
%     C_{yx} & C_{yy}  \\
%    \end{array}
%    \right)
%\end{align}
%So we can obtain:
%\begin{align}
%      \rho = \max_{P_{x},P_{y}} \frac{{P_x}^TC_{xy}{P_y}}{\sqrt{{P_x}^TC_{xx}P_i {P_y}^TC_{yy}P_y }}
%\end{align}
%
%Following \cite{DBLP:journals/neco/HardoonSS04}, this can be optimized by solving a standard eigenproblem like $Ax = \lambda x $. By applying CCA, the original feature of different media types (usually with different dimensions) can be represented by vectors of the same dimension. 
CCA and its variants such as \cite{ Akaho01akernel, ClusterCCA,  ICML2013DCCA, DBLP:journals/ijcv/GongKIL14, DBLP:conf/iccv/RanjanRJ15}, are the most popular baseline methods for cross-media retrieval.

CCA itself is unsupervised and does not use semantic category labels, but researchers have also attempted to extend CCA for incorporating semantic information. Rasiwasia et al. \cite{RasiwasiaMM10SemanticCCA} propose to first apply CCA to get the common space of image and text, and then achieve the semantic abstraction by logistic regression. Costa et al. \cite{costa2014role} then further verify the effectiveness of combining CCA with semantic category labels. GMA \cite{CVPR2012GMA} also obtains the improvement on accuracy, which is a supervised extension of CCA.  Multi-view CCA \cite{DBLP:journals/ijcv/GongKIL14} is proposed to take high-level semantics as the third view of CCA, and multi-label CCA \cite{DBLP:conf/iccv/RanjanRJ15} is designed to deal with the scenarios where cross-media data has multiple labels. These methods achieve considerable progress, which indicates that semantic information is helpful to improve the accuracy of cross-media retrieval.
Besides CCA, there are also alternative methods of traditional statistical correlation analysis. For example, cross-modal factor analysis (CFA) \cite{LiMM03CFA} is proposed to minimize the Frobenius norm between pairwise data in the common space. %In the experiments of \cite{LiMM03CFA}, CFA achieves better accuracy than CCA in most cases since CFA directly models the pairwise correlations in the transformed domain, making the common space more suitable for cross-media retrieval.
As the basic paradigms of cross-media common space learning, these methods are relatively efficient for training and easy to be implemented. However, it is difficult to fully model the complex correlations of cross-media data in the real world only by linear projections. In addition, most of these methods can only model two media types, but cross-media retrieval usually involves more than two media types.
%so they are not sufficient for the practical cross-media retrieval where there are usually more than two media types.
%. But in practical cross-media retrieval problems, the number of media types is often more than two, so when dealing with these problems, the above methods are far insufficient to work well.

\subsection{DNN-based Methods}
\label{section:DnnMethods}
With the great advance of deep learning, deep neural network (DNN) has shown its potential in different multimedia applications such as object recognition \cite{NIPS2013_5204}  and text generation \cite{KirosMNLM2014}. With considerable power of learning non-linear relationships, DNN is also used to perform common space learning for data of different media types. Ngiam et al. \cite{ngiam32011multimodal} apply an extension of restricted Boltzmann machine (RBM) to the common space learning and propose bimodal deep autoencoder, in which inputs of two different media types pass through a shared code layer, in order to learn the cross-media correlations as well as preserve the reconstruction information.
Following this idea, some similar deep architectures are proposed and achieve progress in cross-media retrieval. For example, Srivastava et al. \cite{srivastava42012multimodal} adopt two separate deep Boltzmann machine (DBM) to model the distribution over the features of different media types, and the two models are combined by an additional layer on the top of them as the joint representation layer, which can learn the common space by computing joint distribution. % Similarly, Pang et al. \cite{DBLP:journals/tmm/PangZN15} also adopt DBM for learning highly non-linear relationships among the multimodal inputs from videos, including visual, auditory and textual modalities.

There are also some attempts to combine DNN with CCA as deep canonical correlation analysis (DCCA)  \cite{ICML2013DCCA,DBLP:conf/cvpr/YanM15}. DCCA can be viewed as a non-linear extension of CCA, and used to learn the complex non-linear transformations for two media types. Different from the previous works \cite{ngiam32011multimodal,srivastava42012multimodal} which build one network with a shared layer for different media types, there are two separate subnetworks in DCCA, and the total correlation is maximized by the correlation constraints between the code layers. 
Feng et al.  \cite{feng12014cross} propose three architectures for common space learning: correspondence autoencoder, correspondence cross-modal autoencoder and correspondence full-modal autoencoder. All of them have similar architectures consisting of two subnetworks coupled at the code layers, and jointly consider the reconstruction errors and the correlation loss. Some works also consist of two autoencoders, such as independent component multimodal autoencoder (ICMAE) \cite{zhang2014start} and deep canonically correlated autoencoders (DCCAE) \cite{DBLP:conf/icml/WangALB15}. ICMAE focuses on attribute discovery by learning the shared representation across visual and textual modalities, and DCCAE is optimized by the integration of reconstruction errors and canonical correlations. 
Peng et al. \cite{DBLP:conf/ijcai/PengHQ16} propose cross-media multiple deep networks (CMDN), which is a hierarchical architecture with multiple deep networks. CMDN jointly preserves the intra-media and inter-media information to generate two kinds of complementary separate representations for each media type, and then hierarchically combines them to learn the common space via a stacked learning style, which improves the retrieval accuracy. In addition, in the work of \cite{DBLP:journals/tip/WuLSYZRZ16}, clicks of users are exploited as side information for cross-media common space learning.
A large part of the aforementioned methods are non-convolutional and take hand-crafted features as inputs as \cite{feng12014cross,DBLP:conf/ijcai/PengHQ16}. Wei et al. \cite{DBLP:journals/tcyb/WeiZLWLZY17} propose deep-SM to exploit  convolutional neural network (CNN) with deep semantic matching, which demonstrates the power of the CNN features in cross-media retrieval. He et al. \cite{DBLP:journals/tmm/HeXKWP16} propose a deep and bidirectional representation learning model, with two convolution-based networks modeling the matched and unmatched image/text pairs simultaneously for training.

%Generative Adversarial Networks (GANs) \cite{DBLP:conf/nips/GoodfellowPMXWOCB14} have been increasingly popular, which consist of two models trained simultaneously via an adversarial process, namely a generative model and a discriminative model. Researchers have adopted GANs to conduct image synthesis based on text annotation. Reed et al. \cite{DBLP:conf/icml/ReedAYLSL16} develop a GAN formulation to convert the visual concepts from characters to pixels. Later, they further propose the Generative Adversarial What-Where Network (GAWWN) \cite{DBLP:conf/nips/ReedAMTSL16} to synthesize images by giving location of the content to draw. However, the exploration of GANs on cross-media scenario is still at a preliminary stage and the challenges are existed for effectively modeling the large-scale cross-media data with complex correlations. 

The deep architectures used in cross-media retrieval mainly include two ways. The first way can be viewed as one network, and inputs of different media types pass through the same shared layer \cite{ngiam32011multimodal, srivastava42012multimodal}, while the second way consists of subnetworks coupled by correlation constraints at the code layers \cite{feng12014cross, WangStackedAE2014}. These methods take DNN as the basic model, so have the advantage of abstraction ability for dealing with complex cross-media correlations. However, training data usually plays a key role for the performance of DNN model, and large-scale labeled cross-media datasets are much harder to collect than single-media datasets. 
It is noted that most of the above works have the limitation of taking only two media types as inputs, although there exist some recent works for more than two kinds of inputs such as \cite{DBLP:journals/corr/CastrejonAVPT16} which takes five input types. Jointly learning common space for more than two media types can improve the flexibility of cross-media retrieval, which is an important challenge for future research. 

Except for the above works, other deep architectures have also been designed for multimedia applications such as image/video caption and text to image synthesis \cite{DBLP:journals/corr/MaoXYWY14a, DBLP:conf/cvpr/VinyalsTBE15, DBLP:conf/icml/ReedAYLSL16, DBLP:conf/nips/ReedAMTSL16}.
For example, recurrent neural network (RNN) and long short-term memory (LSTM) \cite{DBLP:journals/corr/MaoXYWY14a, DBLP:conf/cvpr/VinyalsTBE15} have been applied to image/video caption, which can generate text descriptions for visual content. 
%Tang et al. \cite{DBLP:journals/tomccap/TangSLQW16} propose to perform cross-media transfer learning with DNN for propagating knowledge from text to image to improve the performance of image classification. 
Generative adversarial networks (GANs) \cite{DBLP:conf/nips/GoodfellowPMXWOCB14} are proposed by Goodfellow et al., which estimate generative models via an adversarial process by simultaneously training two models: a generative model and a discriminative model.
The basic idea of GANs is to set up a game between two players, and pit two adversaries against each other. Each player is represented by a differentiable function, which are typically implemented as deep neural networks according to \cite{DBLP:journals/corr/Goodfellow17}.
Reed et al. \cite{DBLP:conf/icml/ReedAYLSL16} develop a GANs formulation to convert the visual concepts from characters to pixels. Later, they propose generative adversarial what-where network (GAWWN) \cite{DBLP:conf/nips/ReedAMTSL16} to synthesize images by giving the locations of the content to draw.
These methods are not directly designed for cross-media retrieval, but their ideas and models are valuable to it.

\subsection{Cross-media Graph Regularization Methods}
\label{section:graphReg}
Graph regularization \cite{GraphRe} is widely used in semi-supervised learning, which considers semi-supervised learning problem in the view of labeling a partially labeled graph. The edge weights denote the affinities of data in the graph, and the aim is to predict the labels of unlabeled vertices. Graph regularization can enrich the training set and make the solution smooth.

%Learning common space projection with graph regularization can deal with semi-supervised data, so as to enrich the training set. On the other hand, media objects with different media types can be constructed into the same graph, which is very convenient to handle more than two media types.

%To the best of our knowledge, Zhai et al. \cite{ZhaiAAAI2013JGRHML} make the first attempt to
Zhai et al. \cite{ZhaiAAAI2013JGRHML} propose joint graph regularized heterogeneous metric learning (JGRHML). They incorporate graph regularization into the cross-media retrieval problem, which uses data in the learned metric space for constructing the joint graph regularization term. Then they propose joint representation learning (JRL) method \cite{ZhaiTCSVT2014JRL} with the ability of jointly considering correlations and semantic information in a unified framework for up to five media types. Specifically, they construct a separate graph for each media type, in which the edge weights denote affinities of labeled and unlabeled data of the same media type. With graph regularization, JRL enriches the training set and learns a projection matrix for each media type jointly. 
% This approach learns sparse projection matrix for different types of media simultaneouslypes can align each other and thus be robust for the noise in different types of media. JRL uses a semi-supervised regularization approach to exploit both the labeled data and unlabeled data of all the different types of media, so the unlabeled examples of media increase the diversity of training data and boost the performance of joint representation learning.
%Zhai et al. \cite{ZhaiAAAI2013JGRHML} proposed a joint graph regularized heterogeneous metric learning (JGRHML) algorithm. In JGRHML, different types of media are complementary to each other and optimizing them simultaneously can make the solution smoother for both media. They then further learn an explicit high-level semantic representation through label propagation based on a unified K-nearest neighbor graph, which is constructed from all of the labeled and unlabeled heterogeneous data. Therefore, both heterogeneous similarities and homogeneous similarities are incorporated into the unified graph  \cite{ChungAMS97SpectralGraph}.
As JRL separately constructs different graphs for different media types, Peng et al. \cite{PengHypergraph2015} further propose to construct a unified hypergraph for all media types in the common space, thus different media types can boost each other. Another important improvement of \cite{PengHypergraph2015} is to utilize fine-grained information by media instance segmentation, which helps to exploit multi-level correlations of cross-media data. Graph regularization is also an important part in some recent works such as \cite{DBLP:journals/tmm/WuYYTZZ14, PAMI15Graph, DBLP:conf/sigir/LiangLCHW16}, where a cross-media graph regularization term is used to preserve the intra-media and inter-media similarity relationships. 

Graph regularization is effective for cross-media correlation learning because it can describe various correlations of cross-media data, such as semantic relevance, intra-media similarities and inter-media similarities. Besides, graph regularization can naturally model more than two media types in a unified framework \cite{PengHypergraph2015}. However, the graph construction process usually leads to high time and space complexity, especially in real-world scenarios with large-scale cross-media data.

\subsection{Metric Learning Methods}
Metric learning methods are designed to learn transformations of input features from the given similar/dissimilar information to achieve better metric results, which are widely used in single-media retrieval \cite{QuadriantoMetric2011, ICML2010Rank}. It is natural to view cross-media data as the extension of multi-view single-media data, so researchers attempt to apply metric learning to cross-media retrieval directly. Intuitively, we can learn two transformations for two media types, and let similar instances be close and dissimilar instances be apart \cite{WuMetric2010}.
%, which only uses the similarity and dissimilarity information of cross-media data.
JGRHML \cite{ZhaiAAAI2013JGRHML} is a representative work of cross-media metric learning, which has also been discussed in Section \ref{section:graphReg}. Besides the similar/dissimilar information, JGRHML introduces a joint graph regularization term for the metric learning. Different media types are complementary in the joint graph regularization and optimizing them jointly can make the solution smooth.
%In their latter work of \cite{ZhaiTCSVT2014JRL}, the method is extended as Joint Representation Learning (JRL). Although no longer called Metric Learning, the modeling idea is still the same. The key improvement is the ability for simultaneously modeling more than two media types.
Metric learning preserves the semantic similar/dissimilar information during the common space learning, which is important for semantic retrieval of cross-media data. 
%Different regularization terms can be incorporated into metric learning to deal with complex situations. For example, JGRHML takes the scale regularization term to reduce overfitting, and uses the joint graph regularization term to simultaneously model different media types. 
However, the main limitation of existing metric learning methods for cross-media retrieval such as \cite{WuMetric2010, ZhaiAAAI2013JGRHML} is that they depend on the supervision information, and are not applicable when supervision information is unavailable. 

\subsection{Learning to Rank Methods}
%While most cross-media retrieval methods mainly take the similarity between pairwise data for training, L
Learning to rank methods take the ranking information as training data, and directly optimize the ranking of retrieved results, instead of the similarities between pairwise data.
%Most cross-media retrieval methods mainly focus on the pairwise data similarities, but in practice users will get a ranking list as the retrieval results. Learning to rank aims to directly optimize the rankings of retrieved results, instead of the similarities of pairwise data.
Early works of learning to rank focus on single-media retrieval, but some works such as \cite{IR2010Rank} indicate that they can be extended to cross-language retrieval.
In the work of \cite{GrangierPAMI08TBIR}, a discriminative model is proposed to learn mappings from the image space to the text space, but only uni-directional ranking (text$\rightarrow$image) is involved. For bi-directional ranking methods (specifically text$\rightarrow$image and image$\rightarrow$text ranking information), Wu et al. \cite{WuACMMM2013} propose bi-directional cross-media semantic representation model (Bi-CMSRM) to optimize the bi-directional listwise ranking loss. To incorporate the fine-grained information, Jiang et al. \cite{JiangDeep2015} first project visual objects and text words into the local common space, and then project them into the global common space in a compositional way with ranking information. In addition, Wu et al. \cite{DBLP:journals/tip/WuJLTLZZ15} take a conditional random field for shared topic learning, and then perform latent joint representation learning with ranking function.
%The latent space embedding is discriminatively learned by the structural large margin learning.
Leaning to rank is designed for directly benefiting the final retrieval performance, and can serve as the optimization objective for cross-media retrieval. Existing methods mainly involve only two media types such as \cite{GrangierPAMI08TBIR,WuACMMM2013,DBLP:journals/tip/WuJLTLZZ15}, and when the number of media types increases, there remains a problem on how to incorporate the ranking information of more than two media types into a unified framework.

\subsection{Dictionary Learning Methods}
Dictionary learning methods hold the view that data consists of two parts: dictionaries and sparse coefficients. The idea can also be incorporated into cross-media retrieval: decomposing data into the media-specific part for each media, and the common part for cross-modal correlations.
%So if the mapping of dictionaries is constructed, data can be projected into the space of the other media type, and then the similarity will be measured.
Monaci et al. \cite{TIP07DL} propose to learn the multi-modal dictionaries for recovering meaningful synchronous patterns from audio and visual signals. The key idea of this method is to learn the joint audio-visual dictionaries, so as to find the temporal correlations across different modalities. %However, this is not a cross-media retrieval method, since it can only take the synchronous temporal signals as input.
However, since it only takes the synchronous temporal signals as inputs, it is not a cross-media retrieval method.
Jia et al. \cite{NIPS2010DL} propose to learn one dictionary for each modality, while the weights of these dictionaries are the same. In this work, data is clearly decomposed into two parts: the private dictionaries and the shared coefficients. %However, it is too restricted to assume that the common weights are the same for all the media types.
Zhu et al. \cite{ZhuDL2014} propose cross-modality submodular dictionary learning (CmSDL), which learns a modality-adaptive dictionary pair and an isomorphic space for cross-media representation.

Coupled dictionary learning \cite{CVPR2012DL} is an effective way to jointly construct the private dictionaries for two views. Zhuang et al.\cite{zhuang2013supervised} propose to extend single-media coupled dictionary learning to cross-media retrieval, assuming that there exist linear mappings among sparse coefficients of different media types. Data of one media type can be mapped into the space of another media type via these sparse coefficient mappings. %Moreover, the label information is employed to discover the shared structure inside each media type within the same category by a mixed norm \cite{NieNIPS10L21FeatureSelection, YangIJCAI2011L21Unsepervised}.
In conclusion, dictionary learning methods model cross-media retrieval problem in a factorization way, and the common space is for sparse coefficients. Based on this idea, they have different views of methods, such as a unique sparse coefficient for all media types \cite{NIPS2010DL} and a set of projections among sparse coefficients of different media types \cite{zhuang2013supervised}. It is easier to capture cross-media correlations from the sparse coefficients of different media types due to the high sparsity. However, it is a challenge to solve the optimization problem with mass calculation of dictionary learning on large-scale cross-media data.

%However, it should be noted that some works like \cite{NIPS2010DL} take strong assumptions on the sparse code correlations, which limits the flexibility of dictionary learning in cross-media retrieval.
%As a result, the multi-modal retrieval is conducted via a set of jointly learned mapping functions across multi-modal data.
%Note that there is no such a common space in the general sense in SliM{\textsuperscript 2}. For example, an image will be projected into the text space, not an intermediate space. In this scenario, we can call text space as the common space. This may sound reasonable, but generally a common space should be a low-dimensional semantic space, which is not adequate here.

\subsection{Cross-media Hashing Methods}
Nowadays, the amount of multimedia data is growing dramatically, which requires high efficiency of retrieval system. Hashing methods are designed for accelerating retrieval process, and widely used in various retrieval applications. However, most of them only involve a single media type such as image \cite{DBLP:journals/pieee/WangLKC16}. For example, Tang et al. \cite{DBLP:journals/tip/TangLWZ15} propose to learn image hashing functions with discriminative information of the local neighborhood structure, and exploit the neighbors of samples in the original space to improve the retrieval accuracy.
Cross-media hashing aims to generate the hash codes for more than one media type, and project the cross-media data into a common Hamming space.
%Some works extends the hashing methods such as the similarity-sensitive hashing and the spectral hashing, to the scenarios of multiple input spaces \cite {bronstein2010data, kumar2011learning}. To model the multiple information sources, Zhang et al. \cite{zhang2011composite} propose the composite hashing with multiple information sources, and the idea is to preserve both the similarities in original spaces and the correlations between the information sources. This method is not designed for cross-media retrieval, but the idea is valuable for it, which is shown in some later work \cite{zhen2012co, hu2014iterative, zhen2012probabilistic}.
There have been some works extending single-media hashing to the retrieval of data with multiple views or information sources such as \cite{kumar2011learning, zhang2011composite}. They are not designed specifically for cross-media retrieval, but these methods and ideas can be easily applied to it. For example, to model the multiple information sources, Zhang et al. \cite{zhang2011composite} propose composite hashing with multiple information sources (CHMIS), and the idea is to preserve both the similarities in the original spaces and the correlations between multiple information sources. Similarly, the idea of preserving both inter-media and intra-media similarities is a key principle in some later works of cross-media retrieval \cite{zhen2012co, hu2014iterative, zhen2012probabilistic, DBLP:journals/tmm/WuYYTZZ14}. For instance, Wu et al. \cite{DBLP:journals/tmm/WuYYTZZ14} first apply hypergraph to model intra-media and inter-media similarities, and then learn multi-modal dictionaries for generating hashing codes.
The discriminative capability of hash codes has also been taken into consideration \cite{YuSIGIR2014DCDH}, which helps to learn the hash codes under supervised condition. 
More recently, Long et al. \cite{DBLP:conf/sigir/LongCWY16} propose to learn common space projection and composite quantizers in a seamless scheme, while most existing works view continuous common space learning and binary codes generation as two separate stages.

Besides, cross-media hashing methods are various such as \cite{bronstein2010data,rastegari2013predictable, DingCollectiveHash2014, ZhaiParametricHash2013, zhuang2014cross, DBLP:conf/ijcai/WangSS15, DBLP:conf/ijcai/WangGWH15, DBLP:conf/ijcai/LiuJWH16, DBLP:conf/ijcai/WangCO015,ZhouLatentSemanticHash2014}. % They all aim to learn hashing functions to project heterogeneous data into a common Hamming space. 
The models vary from eigen-decomposition and
boosting \cite{bronstein2010data} to the probabilistic generative model \cite{zhen2012probabilistic}, and even the deep architectures \cite{zhuang2014cross, DBLP:conf/ijcai/WangCO015}. The aforementioned cross-media hashing methods mainly consider similar factors such as inter-media similarities, intra-media similarities and semantic discriminative capability. 
It is noted that cross-media hashing methods are learning-based, because they learn from the cross-media correlations to bridge the ``media gap".
Cross-media hashing has the advantage on retrieval efficiency due to the short binary hashing codes, which can benefit the retrieval on large-scale datasets in the real world. However, existing works such as \cite{zhuang2014cross, DBLP:conf/ijcai/WangSS15, DBLP:conf/ijcai/WangGWH15, DBLP:conf/ijcai/LiuJWH16, DBLP:conf/ijcai/WangCO015,ZhouLatentSemanticHash2014} only involve the retrieval between data of two media types (mainly image and text). As for the experiments, some small-scale datasets such as Wikipedia (2,866 image/text pairs) and Pascal Sentence (1,000 image/text pairs) are used to evaluate the accuracy of hashing \cite{ZhouLatentSemanticHash2014, hu2014iterative, DingCollectiveHash2014, zhuang2014cross, rastegari2013predictable}. Nevertheless, the efficiency advantage of hashing cannot be effectively validated on such small-scale datasets.

\subsection{Other Methods}
There are still some methods that cannot be easily categorized into the aforementioned categories. They also follow the idea to project heterogeneous data into a common space, so that the similarities of them can be directly measured.
For example, Zhang et al. \cite{ZhangSCP2014} propose a two-step method, which first projects the data into a high-dimensional common space, and then maps data from the high-dimensional space into a low-dimensional common space, according to the intra-class distance and inter-class distance. Kang et al. \cite{DBLP:journals/tmm/KangXLXP15} propose local group based consistent feature learning (LGCFL) to deal with unpaired data. In this method, the common space learning can be obtained according to the semantic category labels, instead of strictly paired data as CCA.

Most existing methods only learn one projection for each media type such as \cite{HotelingBiometrika36RelationBetweenTwoVariates,RasiwasiaMM10SemanticCCA,ZhaiTCSVT2014JRL}. This can be further explained as two main aspects: The first is learning only one global projection for each media type, which may lead to inflexibility for large-scale and complex data. Instead, Hua et al. \cite{HuaTina2014} propose to learn a set of local projections, and analyze the hierarchy structure to exploit the semantic correlations of data tags. The second is using the same projections for all retrieval tasks (such as image$\rightarrow$text and text$\rightarrow$image retrieval). Instead, Wei et al. \cite{DBLP:journals/tist/WeiZZWXFY16} propose to learn different projection matrices for image$\rightarrow$text retrieval and text$\rightarrow$image retrieval. The idea of training different models for different tasks is also presented in the works such as \cite{DBLP:conf/bmvc/VermaJ14}. However, there exists a limitation for this idea that as the number of retrieval tasks increases, the number of projection matrices to be learned will also increase. 

Following the attempts of discovering subspaces and manifolds that provide the common low-dimensional representations of two different high-dimensional datasets \cite{ Manifold05} (for example, two image datasets from different viewpoints), some works such as \cite{mao2013parallel} also extend manifold alignment for cross-media retrieval. These methods take the intuition that high-dimensional data has low-dimensional manifold structures, and aim to find the common space projections for different media types by aligning their underlying manifold representations.
% Manifold alignment takes the intuition that the high-dimensional data lie on low-dimensional manifold structures. These geometric structures make it inappropriate to solely use the Euclidean distance to measure the distances in the high-dimensional space. %Note that the methods of manifold alignment mostly involve the construction of KNN graphs, which leads to the high complexity of time and space when dealing with large-scale datasets.

\section{Cross-media Similarity Measurement}
\label{section:similarity}

Cross-media similarity measurement methods aim to measure similarities of  heterogeneous data directly, without explicitly projecting media instances from their separate spaces to a common space.
For the absence of common space, cross-media similarities cannot be computed directly by distance measuring or normal classifiers. An intuitive way is to use the known media instances and correlations in datasets as the basis to bridge the ``media gap".

Existing methods for cross-media similarity measurement usually take the idea of using edges in graphs for representing the relationships among media instances and multimedia documents (MMDs). According to the different focuses of methods, we further classify them into two categories as subsections: \emph{(A) Graph-based methods} focus on the construction of graphs, and \emph{(B) neighbor analysis methods} mainly consider how to exploit the neighbor relationships of data for similarity measurement. These two categories have overlaps in algorithm process, because the neighbor relationships may be analyzed in a constructed graph.  %However, Neighbor Analysis Methods usually don't take the assumption that the MMDs are available so are more flexible.

%Here we mainly introduce two categories of methods, graph-based Methods and Neighbor Analysis Methods. Graph-based methods
%Although they have some overlap in algorithm process, there are still important differences in the view of problem: graph-based methods focus on the construction of graphs which indicate the cross-media correlation, while neighbor analysis methods mainly consider how to use the neighbor relationships of data for cross-media retrieval.

\subsection{Graph-based Methods}

The basic idea of graph-based methods is to view the cross-media data as vertices in one or more graphs, and the edges are constructed by the correlations of cross-media data. Single-media content similarities, co-existence relationships and semantic category labels can be jointly used for graph construction. Through a process such as similarity propagation \cite{ZhuangTMM08SemanticCorrelation} and constraint fusion \cite{tong2005graph}, the retrieval results can be obtained.
These methods often focus on the situations when the relevance in MMDs is available, which contain data with multiple media types of the same semantics \cite{DBLP:journals/pr/YangWXZC10}. The co-existence relationships of data in MMDs provide important hints to bridge different media types. %, which is similar to the methods of multi-modal information fusion like the cross-media topic detection \cite{ChuTD2014}. 
For example, a graph indicating the similarities of MMDs plays an important role in \cite{YangMM09LRGA, YangIEEETPAMI2012}, and the cross-media retrieval is based on MMD affinities in this graph.

Tong et al. \cite{tong2005graph} construct an independent graph for each media type. These graphs are further combined by linear fusion or sequential fusion, and then the similarity measurement of cross-media data is conducted. Different from \cite{tong2005graph}, Zhuang et al. \cite{ZhuangTMM08SemanticCorrelation} construct a uniform cross-media correlation graph, which integrates all media types. The edge weights are determined by the similarities of single-media data and co-existence relationships. Besides, the links among MMDs on web pages have also been taken into account in the work of \cite{zhuang2006approach}. Yang et al. \cite{YangTMM08MMDManifolds} propose a two-level graph construction strategy. They first construct two types of graphs: one graph for each media type, and the other for all MMDs. Then the characteristics of media instances propagate along the MMD semantic graph and the MMD semantic space is constructed to perform cross-media retrieval. While existing methods mostly consider only the positive correlations in similarity propagation, Zhai et al. \cite{ZhaiICASSP2012CCP} propose to propagate both the positive and negative correlations among data of different media types in graphs, and improve the retrieval accuracy.

The key idea of graph-based similarity measurement methods is to construct one or more graphs, and represent cross-media correlations on the level of media instances or MMDs. %The graph construction strategy can be similar to Cross-media Graph Regularization Methods in Section III. B, but here the cross-media similarity is obtained directly with the graph by methods as similarity propagation, instead of acting as the regularization term for common space learning.
It is helpful to incorporate various types of correlation information by graph construction. However, graph-based methods are time and space consuming due to the process of graph construction. In addition, existing works are often devoted to scenarios where the relevance in MMDs is available, and relevance feedback usually acts as a key factor in these works such as \cite{YangMM09LRGA, ZhuangTMM08SemanticCorrelation, YangIEEETPAMI2012}. On the one hand, when the above relevance is unavailable, it would be difficult to perform cross-media retrieval, especially when the queries are out of the datasets. On the other hand, in real-world applications, the relationships of MMDs are usually noisy and incomplete, which is also a key challenge for these methods.

\subsection{Neighbor Analysis Methods}

Generally speaking, neighbor analysis methods are usually based on graph construction because the neighbors may be analyzed in a given graph \cite{ZhuangTMM08SemanticCorrelation, YangTMM08MMDManifolds}. In this paper, graph-based methods mainly involve the process of graph construction, while neighbor analysis methods focus on using the neighbor relationships for similarity measurement.

Clinchant et al. \cite{clinchant2011semantic} introduce a multimedia fusion strategy named transmedia fusion for cross-media retrieval. For instance, there exists a dataset containing image/text pairs, and users retrieve the relevant texts by queries of images. Given one query of image, its nearest neighbors will be retrieved according to single-media content similarities, and then the text descriptions of these nearest neighbors are regarded as the relevant texts. Zhai et al. \cite{ZhaiMMM2012HSNN} propose to compute cross-media similarities by the probabilities of two media instances belonging to the same semantic category, which are calculated by analyzing the homogeneous nearest neighbors of each media instance. Ma et al. \cite{MaICIP2013CBCA} propose to compute cross-media similarities in the perspective of clusters. In their work, clustering algorithm is first applied to each media type, and then the similarities among clusters are obtained according to the data co-existence relationships. The queries will be assigned to clusters with different weights according to the single-media content similarities, and then retrieval results can be obtained by computing the similarities among clusters.

Neighbor analysis methods find the nearest neighbors in datasets with the queries to get the retrieval results. These neighbors can be used as expanded queries, and serve as the bridges for dealing with queries out of the datasets. In addition, some methods such as \cite{ZhaiMMM2012HSNN} do not rely on MMDs, so they are flexible. However, because the neighbor analysis methods may be actually based on graph construction, they have the same problem of high time and space complexity. It is also difficult to ensure the relevant relationships of neighbors, so the performance is not stable.

\section{Other Methods for Cross-media Retrieval}
\label{section:other}
Besides common space learning and cross-media similarity measurement methods, we introduce two categories of other cross-media methods as subsections: \emph{(A) Relevant feedback analysis} is an auxiliary method for providing more information on user intent to promote the performance of retrieval. \emph{(B) Multimodal topic model} views cross-media data in the topic level, and the cross-media similarities are usually obtained by computing the conditional probability. 

\subsection {Relevance Feedback Analysis}
To bridge the vast ``media gap", the relevance feedback (RF) is beneficial to provide more accurate information and facilitate the retrieval accuracy. It is worth noting that RF is widely used in cross-media similarity measurement, and the effectiveness has been validated in some works \cite{YangMM09LRGA, YangIEEETPAMI2012,YangTMM08MMDManifolds}.
RF includes two types: short-term feedback and long-term feedback. Short-term feedback only involves RF information provided by the current user, while long-term feedback takes RF information provided by all users into account. For short-term feedback, in the works of \cite{YangMM09LRGA, YangTMM08MMDManifolds}, when the queries are out of the datasets, the system will show the nearest neighbors in the dataset with the queries, and users should label them as the positive or negative samples. Then the similarities will be refined according to the feedback. For long-term feedback, Yang et al. \cite{YangIEEETPAMI2012} propose to convert long-term feedback information into pairwise similar/dissimilar constraints to refine the vector representation of data. Zhuang et al. \cite{ZhuangTMM08SemanticCorrelation} exploit both long-term and short-term feedback. As for long-term feedback, they first investigate the global structure of all feedback and then refine the uniform cross-media correlation graph. For short-term feedback, they simply use the positive samples as expanded queries. RF is an auxiliary technique to improve the retrieval accuracy in an interactive way, but with the cost of human labor.

%\subsection{Metric Learning}

\subsection{Multimodal Topic Model}
Inspired by topic models such as latent dirichlet allocation (LDA) \cite{LDA} in text processing, researchers have extended topic models to the multimodal retrieval.
%The use of LDA in cross-media retrieval is pretty flexible, but these methods face the same problem: how to model the topic correlation between different media types?
These models are often designed for applications such as image annotation, involving images and their corresponding tags. %Compared to cross-media retrieval which emphasizes the flexibility of media types, such problems as \cite{JeonSIGIR03ImageAnnotation} are mostly limited to modeling images and the text. Additionally, image annotation mainly determines the probability of a single tag being assigned to an image, while in cross-media retrieval, ``text" refers to the text descriptions which contain the rich semantic information, rather than only the tags.
Correspondence LDA (Corr-LDA) \cite{cLDA} is a classical multimodal extension of LDA for image annotation. Specifically, it first generates image region descriptions and then generates the caption. % as shown in Fig \ref{fig:clda}.
% The model is depicted in Fig \ref{fig:clda}.
However, it takes a strict assumption that each image topic must have a corresponding text topic. To address this problem, topic-regression multi-modal LDA (tr-mmLDA) \cite{TrmmLDA} uses two separate topic models for image and text, and finally applies a regression module to correlate the two hidden topic sets. Nevertheless, it still takes a strong assumption that each word in the text has a visual interpretation.
%\begin{figure}[!t]
%\begin{center}
%   \includegraphics[width=0.8\linewidth]{Fig/cLDA.eps}
%\end{center}
%   \caption{A brief illustration for Corr-LDA model. It first generates {\em N} image region descriptions {\em r}, and then generates the caption words for the image regions.}
   % from {\em M} caption words.}
%\label{fig:clda}
%\end{figure}
To further make the topic models flexible and perform cross-media retrieval, Jia et al. \cite{JiaICCV2011MDRF} propose multi-modal document random field (MDRF) method, which can be viewed as a Markov random field over LDA topic models. Wang et al. \cite{wang22014multi} propose a downstream supervised topic model, and build a joint cross-modal probabilistic graphical model to discover the mutually consistent semantic topics.
%In their approach semantic latent topics is generated by multi-modal data in the same multi-modal documents, so the co-occurrence relationship should be relied on here.
Multimodal topic model aims to analyze the cross-media correlations in the topic level. However, these existing methods often take strong assumptions on the distribution of cross-media topics, such as the existence of the same topic proportions or pairwise topic correspondences between different media types, which are not satisfied in real-world application.

%\subsection{Learning to Rank}

%\subsection{Dictionary Learning}

%\begin{figure}[t]
%\begin{center}
%   \includegraphics[width=0.9\linewidth]{Fig/WikiExample.eps}
%\end{center}
%   \caption{Examples from 3 categories of Wikipedia dataset.}
%\label{fig:exmWik}
%\end{figure}

%\begin{figure}[t]
%\begin{center}
%   \includegraphics[width=\linewidth]{Fig/PubMedExample.eps}
%\end{center}
 %  \caption{Examples from 3 categories of XMedia dataset.}
%\label{fig:exmPub1}
%\end{figure}

\section{Cross-Media Retrieval Dataset}
\label{section:dataset}

% Table generated by Excel2LaTeX from sheet '实验结果AVE'

\begin{table*}[htbp]
  \centering
  \caption{The using frequencies of several popular widely-used datasets in the references on cross-media retrieval.}
    \begin{tabular}{c|c|c|c|c|c|c|c|c}
    \hline
       Dataset   & Wikipedia & NUS-WIDE & Pascal VOC 2007 & Pascal Sentence & MIR-Flickr & Corel & LabelMe & XMedia \\
   \hline\hline
    Frequency & 38    & 30    & 9  & 9  & 7     & 6     & 5   &  3   \\
    \hline
    \end{tabular}%
 \label{tab:datasetsFre}%
 \vspace{-2mm}
\end{table*}

Datasets are important for the evaluation of cross-media retrieval methods. We study all references of this paper and summarize the frequencies of several popular datasets in Table \ref{tab:datasetsFre}.
It is shown that Wikipedia and NUS-WIDE datasets are the most widely-used cross-media retrieval datasets. Pascal VOC datasets are a series of important datasets for cross-media retrieval, and also the basis of Pascal Sentence dataset. Pascal VOC 2007 dataset is the most popular one among Pascal VOC datasets. In addition, XMedia dataset is the first cross-media dataset that contains up to five media types. %So we choose these four datasets (Wikipedia, XMedia, NUS-WIDE and Pascal VOC 2007) for evaluation. 
We first introduce Wikipedia and XMedia datasets which are specifically designed for cross-media retrieval, then the rest NUS-WIDE and Pascal VOC 2007 datasets. Besides, we also introduce a large-scale click-based dataset Clickture. %
% with 40 million images and 73.6 million text queries.
%, 11.7 million queries and the click data as the correlation between them.
%We can see that Wikipedia, NUS-WIDE, and Pascal VOC 2007 datasets are widely-used cross-media datasets. XMedia dataset is the first cross-media dataset that contains 5 types of media. So we choose these four datasets to evaluate cross-media methods. We will first introduce Wikipedia and XMedia datasets which are specially designed for cross-media retrieval, and then the rest datasets NUS-WIDE and Pascal VOC 2007.

%\begin{table }[tbp]
%  \centering
%  \caption{The using frequency of several popular widely-used datasets in the references of this %survey on cross-media retrieval.}
%    \begin{tabular}{c|c|c|c|c|c|c|c|c}
%    \hline
%      Dataset    & Wikipedia & NUS-WIDE & Pascal VOC 2007 & XMedia \\
%    \hline\hline
%    Frequency & 35    & 26    & 8  & 3 \\
%    \hline
    %Proportion & 36.3\% & 27.5\% & 6.6\%  & 1.1\%  & 6.6\% & 6.6\% & 4.4\% & 3.3\% \\
    %\hline
%    \end{tabular}%
%  \label{tab:datasetsFre}%
%\end{table*}%

\emph{ \textbf{Wikipedia Dataset}}. Wikipedia dataset \cite{RasiwasiaMM10SemanticCCA} is the most widely-used dataset for cross-media retrieval. It is based on ``featured articles" in Wikipedia, which is a continually updated article collection. There are totally 29 categories in ``featured articles", but only 10 most populated categories are actually considered. Each article is split into several sections according to its section headings, and this dataset is finally generated as a set of 2,866 image/text pairs. %Some examples of the dataset are shown in Figure \ref{fig:exmWik}.
Wikipedia dataset is an important benchmark dataset for cross-media retrieval. However, this dataset is small-scale and only involves two media types (image and text). The categories in this dataset are of high-level semantics to be difficultly distinguished, such as warfare and history, leading to confusions for retrieval evaluation. On the one hand, there are some semantic overlaps among these categories. For example, a war (should belong to  warfare category) is usually also a historical event (should belong to history category). On the other hand, even data belonging to the same category may differ greatly on semantics from each other. %
%, which brings noise to the dataset. 

%\item {\bf XMedia dataset}.
\emph{ \textbf{XMedia Dataset}}. For comprehensive and fair evaluation, we have constructed a new cross-media dataset
XMedia. We choose 20 categories such as insect, bird, wind, dog, tiger, explosion and elephant. These categories are specific objects that can avoid the confusions and overlaps. For each category, we collect the data of five media types: 250 texts, 250 images, 25 videos, 50 audio clips and 25 3D models, so there are 600 media instances for each category and the total number of media instances is 12,000. All of the media instances are crawled from famous websites: Wikipedia, Flickr, YouTube, 3D Warehouse and Princeton 3D model search engine.
%Each text is a paragraph of an article about the category in Wikipedia and most of the texts are less than 200 words. The images are pictures with high resolution, which contain the objects of each category. Long videos from YouTube are segmented into short clips which exactly represent the categories, and the video clips in this dataset are mostly less than one minute. The collected audio clips are mostly shorter than one minute, which can stand for the categories like wolf howl. The 3D models are objects standing for the 20 semantic categories, such as easily recognizable models like dog and tiger.
%Each text is a paragraph of an article about the category in Wikipedia and most of the texts are less than 200 words. The images are pictures with high resolution, which contain the objects of each category. Long videos from YouTube are segmented into short clips which exactly represent the categories, and the video clips in this dataset are mostly less than one minute. The collected audio clips are mostly shorter than one minute, which can stand for the categories like wolf howl. The 3D models are objects standing for the 20 semantic categories, such as easily recognizable models like dog and tiger.
%Some examples of the dataset are shown in Figure \ref{fig:exmPub1}.
XMedia dataset is the first cross-media dataset with up to five media types (text, image, video, audio and 3D model), and has been used in our works \cite{ZhaiAAAI2013JGRHML,ZhaiTCSVT2014JRL,PengHypergraph2015} to evaluate the effectiveness of cross-media retrieval. XMedia dataset is publicly available and can be accessed through the link: \href{http://www.icst.pku.edu.cn/mipl/XMedia}{http://www.icst.pku.edu.cn/mipl/XMedia}.

% and the correlation information provided by XMedia is solely the semantic categories.
%It should be also noted that there are no pairs of media instances in XMedia dataset like Wikipedia dataset, so the correlation information solely depends on the semantic categories. Some cross-media methods like CCA is based on pairwise correlation, and they may have difficulty using XMedia dataset.

\emph{ \textbf{NUS-WIDE Dataset}}. NUS-WIDE dataset \cite{chua2009nus} is a web image dataset including images and their associated tags. The images and tags are all randomly crawled from Flickr through its public API. With the duplicated images removed, there are 269,648 images in NUS-WIDE dataset of 81 concepts. Totally 425,059 unique tags are originally associated with these images. However, to further improve the quality of tags, those tags that appear no more than 100 times and do not exist in WordNet \cite{DBLP:journals/cacm/Miller95} are removed. So finally 5,018 unique tags are included in this dataset. %NUS-WIDE is a relatively large dataset, but it only contains images and their tags.

\emph{ \textbf{Pascal VOC 2007 Dataset}}. Pascal visual object classes (VOC) challenge \cite{DBLP:journals/ijcv/EveringhamGWWZ10} is a benchmark in visual object category detection and recognition. Pascal VOC 2007 is the most popular Pascal VOC dataset, which consists of 9,963 images divided into 20 categories. The image annotations serve as the text for cross-media retrieval, and are defined over a vocabulary of 804 keywords. % This dataset is divided into a collection of 5,011 images as the training set, and a collection of 4,952 images as the testing set.

\emph{ \textbf{Clickture Dataset}}. Clickture dataset \cite{DBLP:conf/mm/HuaYWWYWRL13} is a large-scale click-based image dataset, which is collected from one-year click-through data of a commercial image search engine. The full Clickture dataset consists of 40 million images and 73.6 million text queries. It also has a subset Clickture-Lite with 1.0 million images and 11.7 million text queries. Following recent works as \cite{DBLP:conf/mm/PanYTLN14, DBLP:journals/tip/WuLSYZRZ16}, we take Clickture-Lite for experimental evaluation. The training set consists of 23.1 million query-image-click triads, where ``click" is an integer indicating the relevance between the image and query, and the testing set has 79,926 query-image pairs generated from 1,000 text queries. 
%There are 11.7 million distinct queries and 1.0 million unique images in the training set. For the development set, there are 79,926 query-image pairs generated from 1,000 queries provided for testing.

%The training set consists of 23.1 million query-image-click triads, where each query is a textual word/phrase, image is a base64-encoded thumbnail, and click is an integer. There are 11.7 million distinct queries and 1.0 million unique images in the training set. For the development set, there are 79,926 query-image pairs generated from 1,000 queries provided for testing.

Among the above datasets, Wikipedia and XMedia datasets are specifically designed for cross-media retrieval. NUS-WIDE and Pascal VOC 2007 datasets are image/tag datasets, which are initially designed for the evaluation of other applications such as image annotation and classification. There are only tags in these two datasets as text, instead of sentences or paragraph descriptions as Wikipedia and XMedia datasets. Clickture dataset is the largest among these datasets, but it provides no category labels as supervision information.
%There are only tags in the two datasets as text, instead of text description as sentences or paragraphs in Wikipedia and XMedia datasets, and researchers often treat the tag vectors as raw text feature.

\section{Experiments}
\label{section:experiment}

\subsection{Feature Extraction and Dataset Split}

%For fair comparison, we first take exactly the same feature (including the types and dimensions of feature) for each method on cross-media retrieval.
This subsection presents the feature extraction strategy and split of training/testing set in the experiments. 
For Wikipedia, XMedia and Clickture datasets, we take the same strategy as \cite{RasiwasiaMM10SemanticCCA} to generate both text and image representations, and the representations of video, audio and 3D model are the same as \cite{ZhaiTCSVT2014JRL}. In detail, texts are represented by the histograms of a 10-topic LDA model, and images are represented by the  bag-of-visual-words (BoVW) histograms of a SIFT codebook with 128 codewords. Videos are segmented into several video shots first, and then the 128-dimensional BoVW histogram features are extracted for video keyframes. Audio clips are represented by the 29-dimensional MFCC features, and 3D models are represented by the concatenated 4,700-dimensional vectors of a LightField descriptor set. 
For NUS-WIDE dataset, we use the 1,000-dimensional word frequency features and for texts, and the 500-dimensional BoVW features for images provided by Chua et al. \cite{chua2009nus}.
For Pascal VOC 2007 dataset, we use publicly available features for the experiments, which is the same as \cite{DBLP:journals/tmm/KangXLXP15} where the 399-dimensional word frequency features are used for texts, and the 512-dimensional GIST features are used for images. The above feature extraction strategy is adopted for all compared methods in the experiments except DCMIT \cite{DBLP:conf/cvpr/YanM15}, because its architecture contains networks taking the original images and texts as inputs. However, DCMIT does not involve corresponding networks for video, audio and 3D model, so for these 3 media types we use the same extracted features as all the other compared methods.

For Wikipedia dataset, 2,173 image/text pairs are used for training and 693 image/text pairs are used for testing. For XMedia dataset, the ratio of training and testing sets is 4:1 for all the five media types, so we have a training set of 9,600 instances and a testing set of 2,400 instances.
%and the split of XMedia dataset is as  \ref{table:splitXM}. 
For NUS-WIDE dataset, %some of the URLs provided by the dataset are invalid. So 
we select image/text pairs that exclusively belong to one of  the 10 largest categories from valid URLs. As a result, the size of training set is 58,620 and the testing set has a total size of 38,955. %classes, which is strictly the same as \cite{zhuang2013supervised}, and 
Pascal VOC 2007 dataset is split into a training set with 5,011 image/text pairs and a testing set with 4,952 image/text pairs. %Some of the images are multi-labeled, so the images containing only one object are selected in the experiment. 
Images with only one object are selected for the experiments and finally there are 2,808 image/text pairs in the training set and 2,841 image/text pairs in the testing set. For Clickture dataset, there are 11.7 million distinct queries and 1.0 million unique images for training, and 79,926 query-image pairs generated from 1,000 queries for testing.

%\begin{table}[tbp]
%\caption{The split of each type of media on XMedia dataset.}
%\label{table:splitXM}
%\centering{
%\begin{tabular}{c|c|c|c|c|c}
%\hline
%Media  &Image  &Text    &Audio    &Video    &3D  \\ \hline \hline
%Training  &4,000   &4,000   &800  &400  &400\\
%\hline
%Testing   &1,000   &1,000   &200  &100  &100\\ \hline
%\end{tabular}
%}
%\end{table}

% Table generated by Excel2LaTeX from sheet '实验结果分类'
\begin{table*}[htbp]
  \tiny
  \centering
  \caption{The MAP scores of multi-modality cross-media retrieval.}
  \label{table:unifiedretrieval}
    \begin{tabular}{c|c|c|c|c|c|c|c|c|c|c|c|c|c|c}
    \hline
    Dataset & Task  & BITR  & CCA   & CCA+SMN & CFA   & CMCP  & DCMIT  & HSNN  & JGRHML & JRL   & LGCFL & ml-CCA & mv-CCA & {S{\textsuperscript 2}UPG} \\
    \hline \hline
    \multirow{2}{*}{\begin{tabular}{c}Wikipedia\end{tabular}} & {Image$\rightarrow$All} & 0.167  & 0.191  & 0.211  & 0.185  & 0.255  & 0.221  & 0.245  & 0.263  & 0.273  & 0.213  & 0.209  & 0.200  & 0.274  \\
          & {Text$\rightarrow$All} & 0.258  & 0.381  & 0.414  & 0.400  & 0.462  & 0.408  & 0.434  & 0.468  & 0.476  & 0.441  & 0.418  & 0.444  & 0.503  \\
          \hline \hline
    \multirow{5}{*}{\begin{tabular}{c}XMedia\end{tabular}} & {Image$\rightarrow$All} & 0.079  & 0.130  & 0.143  & 0.137  & 0.218  & 0.261  & 0.184  & 0.169  & 0.252  & 0.063  & 0.139  & 0.131  & 0.311  \\
          & {Text$\rightarrow$All} & 0.065  & 0.130  & 0.141  & 0.136  & 0.196  & 0.146  & 0.178  & 0.144  & 0.199  & 0.068  & 0.142  & 0.152  & 0.254  \\
          & {Video$\rightarrow$All} & 0.071  & 0.079  & 0.116  & 0.107  & 0.153  & 0.078  & 0.127  & 0.134  & 0.152  & 0.078  & 0.111  & 0.082  & 0.190  \\
          & {Audio$\rightarrow$All} & 0.072  & 0.104  & 0.125  & 0.117  & 0.202  & 0.100  & 0.187  & 0.131  & 0.197  & 0.078  & 0.133  & 0.105  & 0.227  \\
          & {3D$\rightarrow$All} & 0.061  & 0.070  & 0.082  & 0.111  & 0.219  & 0.070  & 0.160  & 0.203  & 0.181  & 0.086  & 0.154  & 0.069  & 0.291  \\
          \hline \hline
              \multirow{2}{*}{\begin{tabular}{c}NUS-WIDE\end{tabular}} & {Image$\rightarrow$All} & 0.247  & 0.288  & 0.310  & 0.288  & 0.328  & 0.307  & 0.361  & 0.314  & 0.409  & 0.355  & 0.322  & 0.301  & 0.473  \\
                    & {Text$\rightarrow$All} & 0.248  & 0.330  & 0.364  & 0.329  & 0.449  & 0.380  & 0.433  & 0.419  & 0.485  & 0.422  & 0.349  & 0.419  & 0.584  \\
                    \hline \hline
              \multirow{2}{*}{\begin{tabular}{c}Pascal VOC 2007\end{tabular}} & {Image$\rightarrow$All} & 0.088  & 0.116  & 0.179  & 0.150  & 0.258  & 0.185  & 0.273  & 0.150  & 0.329  & 0.270  & 0.265  & 0.194  & 0.345  \\
                    & {Text$\rightarrow$All} & 0.091  & 0.239  & 0.405  & 0.191  & 0.387  & 0.283  & 0.407  & 0.282  & 0.594  & 0.664  & 0.621  & 0.618  & 0.631  \\
                    \hline \hline
    \multicolumn{2}{c|}{Average} & 0.132  & 0.187  & 0.226  & 0.196  & 0.284  & 0.222  & 0.272  & 0.243  & 0.322  & 0.249  & 0.260  & 0.247  & 0.371  \\
    \hline
    \end{tabular}%
\end{table*}%

% Table generated by Excel2LaTeX from sheet '实验结果分类'
\begin{table*}[htbp]
  \tiny
  \centering
  \caption{The MAP scores of bi-modality cross-media retrieval.}
  \label{table:crossretireval}
    \begin{tabular}{c|c|c|c|c|c|c|c|c|c|c|c|c|c|c}
    \hline
    Dataset & Task  & BITR  & CCA   & CCA+SMN & CFA   & CMCP  & DCMIT  & HSNN  & JGRHML & JRL   & LGCFL & ml-CCA & mv-CCA & {S{\textsuperscript 2}UPG} \\
    \hline \hline
    \multirow{2}{*}{\begin{tabular}{c}Wikipedia\end{tabular}} & Image$\rightarrow$Text & 0.222  & 0.249  & 0.277  & 0.246  & 0.326  & 0.277  & 0.321  & 0.329  & 0.339  & 0.274  & 0.269  & 0.271  & 0.377  \\
          & Text$\rightarrow$Image & 0.171  & 0.196  & 0.226  & 0.195  & 0.251  & 0.250  & 0.251  & 0.256  & 0.250  & 0.224  & 0.211  & 0.209  & 0.286  \\
	\hline \hline
    \multirow{20}{*}{\begin{tabular}{c}XMedia\end{tabular}} & Image$\rightarrow$Text & 0.070  & 0.119  & 0.141  & 0.127  & 0.201  & 0.150  & 0.176  & 0.176  & 0.195  & 0.089  & 0.134  & 0.149  & 0.270  \\
          & Image$\rightarrow$Video & 0.089  & 0.098  & 0.197  & 0.174  & 0.203  & 0.092  & 0.182  & 0.239  & 0.225  & 0.097  & 0.137  & 0.158  & 0.264  \\
          & Image$\rightarrow$Audio & 0.085  & 0.103  & 0.162  & 0.129  & 0.219  & 0.087  & 0.225  & 0.198  & 0.234  & 0.121  & 0.148  & 0.135  & 0.265  \\
          & Image$\rightarrow$3D & 0.108  & 0.117  & 0.299  & 0.159  & 0.338  & 0.092  & 0.242  & 0.347  & 0.300  & 0.148  & 0.188  & 0.165  & 0.394  \\
          & Text$\rightarrow$Image & 0.061  & 0.114  & 0.138  & 0.126  & 0.217  & 0.195  & 0.185  & 0.190  & 0.213  & 0.081  & 0.134  & 0.137  & 0.275  \\
          & Text$\rightarrow$Video & 0.081  & 0.110  & 0.127  & 0.137  & 0.174  & 0.110  & 0.158  & 0.201  & 0.160  & 0.074  & 0.121  & 0.118  & 0.242  \\
          & Text$\rightarrow$Audio & 0.079  & 0.127  & 0.155  & 0.136  & 0.213  & 0.127  & 0.215  & 0.183  & 0.209  & 0.083  & 0.160  & 0.155  & 0.242  \\
          & Text$\rightarrow$3D & 0.101  & 0.160  & 0.187  & 0.180  & 0.319  & 0.091  & 0.243  & 0.279  & 0.250  & 0.112  & 0.214  & 0.115  & 0.338  \\
          & Video$\rightarrow$Image & 0.059  & 0.065  & 0.157  & 0.126  & 0.188  & 0.058  & 0.143  & 0.192  & 0.196  & 0.082  & 0.120  & 0.112  & 0.225  \\
          & Video$\rightarrow$Text & 0.065  & 0.078  & 0.090  & 0.102  & 0.141  & 0.078  & 0.118  & 0.134  & 0.123  & 0.072  & 0.090  & 0.090  & 0.193  \\
          & Video$\rightarrow$Audio & 0.074  & 0.093  & 0.137  & 0.117  & 0.155  & 0.093  & 0.166  & 0.139  & 0.152  & 0.110  & 0.128  & 0.096  & 0.168  \\
          & Video$\rightarrow$3D & 0.106  & 0.134  & 0.101  & 0.178  & 0.238  & 0.098  & 0.176  & 0.253  & 0.195  & 0.130  & 0.173  & 0.156  & 0.267  \\
          & Audio$\rightarrow$Image & 0.057  & 0.078  & 0.129  & 0.109  & 0.233  & 0.065  & 0.213  & 0.177  & 0.243  & 0.091  & 0.123  & 0.104  & 0.274  \\
          & Audio$\rightarrow$Text & 0.061  & 0.106  & 0.122  & 0.110  & 0.195  & 0.106  & 0.182  & 0.142  & 0.173  & 0.100  & 0.125  & 0.122  & 0.244  \\
          & Audio$\rightarrow$Video & 0.096  & 0.114  & 0.142  & 0.151  & 0.172  & 0.114  & 0.169  & 0.181  & 0.158  & 0.100  & 0.136  & 0.096  & 0.207  \\
          & Audio$\rightarrow$3D & 0.118  & 0.164  & 0.269  & 0.147  & 0.343  & 0.102  & 0.284  & 0.318  & 0.222  & 0.170  & 0.256  & 0.153  & 0.363  \\
          & 3D$\rightarrow$Image & 0.057  & 0.073  & 0.248 & 0.115  & 0.264  & 0.057  & 0.169  & 0.296  & 0.230  & 0.089  & 0.144  & 0.083  & 0.345  \\
          & 3D$\rightarrow$Text & 0.048  & 0.104  & 0.135  & 0.109  & 0.198  & 0.062  & 0.161  & 0.183  & 0.165  & 0.089  & 0.147  & 0.075  & 0.275  \\
          & 3D$\rightarrow$Video & 0.097  & 0.123  & 0.101  & 0.186  & 0.199  & 0.097  & 0.147  & 0.242  & 0.151  & 0.165  & 0.166  & 0.126  & 0.276  \\
          & 3D$\rightarrow$Audio & 0.080  & 0.153  & 0.214  & 0.147  & 0.255  & 0.085  & 0.259  & 0.318  & 0.185  & 0.101  & 0.205  & 0.106  & 0.329  \\
	\hline \hline
    \multirow{2}{*}{\begin{tabular}{c}NUS-WIDE\end{tabular}} & Image$\rightarrow$Text & 0.279  & 0.250  & 0.365  & 0.251  & 0.368  & 0.330  & 0.404  & 0.451  & 0.466  & 0.388  & 0.345  & 0.345  & 0.525  \\
          & Text$\rightarrow$Image & 0.237  & 0.227  & 0.347  & 0.247  & 0.327  & 0.309  & 0.332  & 0.434  & 0.383  & 0.362  & 0.317  & 0.308  & 0.477  \\
	\hline \hline
    \multirow{2}{*}{\begin{tabular}{c}Pascal VOC 2007\end{tabular}} & Image$\rightarrow$Text & 0.091  & 0.134  & 0.254  & 0.213  & 0.300  & 0.236  & 0.323  & 0.172  & 0.397  & 0.398  & 0.381  & 0.341  & 0.423  \\
          & Text$\rightarrow$Image & 0.095  & 0.123  & 0.201  & 0.196  & 0.265  & 0.249  & 0.266  & 0.170  & 0.263  & 0.325  & 0.303  & 0.261  & 0.326  \\
	\hline \hline
    \multicolumn{2}{c|}{Average} & 0.103  & 0.131  & 0.189  & 0.158  & 0.242  & 0.139  & 0.220  & 0.238  & 0.234  & 0.157  & 0.188  & 0.161  & 0.303  \\
    \hline
    \end{tabular}%
\end{table*}%

\subsection{Evaluation Metrics and Compared Methods}
Two retrieval tasks are conducted for objective evaluation on cross-media retrieval:
\begin{itemize}
\item {\bf Multi-modality cross-media retrieval}. By submitting a query example of any media type, all media types will be retrieved.

\item {\bf Bi-modality cross-media retrieval}. By submitting a query example of any media type, the other media type will be retrieved.

\end{itemize}

On all datasets except for Clickture dataset, both the tasks are performed, and the retrieval results are evaluated by mean average precision (MAP) scores, which are widely adopted in information retrieval. MAP score for a set of queries is the mean of the average precision (AP) for each query. Besides, we also adopt precision-recall curves (PR curves) and running time for comprehensive evaluations. Due to the length limitation of this paper, we  show the PR curves and running time on our website: \href{http://www.icst.pku.edu.cn/mipl/XMedia}{http://www.icst.pku.edu.cn/mipl/XMedia}. 
Clickture dataset does not provide category labels for evaluation with MAP and PR curves. Instead, it consists of many text queries, and for each text query there are multiple images along with the relevance between images and the query, which are uni-directional ground truth. Following \cite{DBLP:conf/mm/PanYTLN14, DBLP:journals/tip/WuLSYZRZ16}, we conduct the text-based image retrieval task for each text query, and take discounted cumulative gain for top 25 results (DCG@25) as evaluation metric.
%Clickture dataset only has groundtruth for Text$\rightarrow$Image retrieval and there is no category information, so the above metric is not applicable. Following \cite{DBLP:conf/mm/PanYTLN14, DBLP:journals/tip/WuLSYZRZ16}, we use text query to retrieve images, and evaluate by Normalized Discounted Cumulative Gain for top 25 results (NDCG@25). 
The compared methods in the experiments include: BITR \cite{DBLP:conf/bmvc/VermaJ14}, CCA \cite{HotelingBiometrika36RelationBetweenTwoVariates}, CCA+SMN \cite{RasiwasiaMM10SemanticCCA}, CFA \cite{LiMM03CFA}, CMCP \cite{ZhaiICASSP2012CCP}, DCMIT \cite{DBLP:conf/cvpr/YanM15}, HSNN \cite{ZhaiMMM2012HSNN}, JGRHML \cite{ZhaiAAAI2013JGRHML}, JRL  \cite{ZhaiTCSVT2014JRL}, LGCFL \cite{DBLP:journals/tmm/KangXLXP15}, ml-CCA \cite{DBLP:conf/iccv/RanjanRJ15}, mv-CCA \cite{DBLP:journals/ijcv/GongKIL14} and S{\textsuperscript 2}UPG  \cite{PengHypergraph2015}. All these methods are evaluated on Wikipedia, XMedia, NUS-WIDE and Pascal VOC 2007 datasets. However, because Clickture dataset provides no category labels for supervised training, only unsupervised methods (BITR, CCA, CFA, DCMIT) are evaluated on this dataset.

\subsection{Experimental Results}
\label{section:Result}
Table \ref{table:unifiedretrieval} shows MAP scores of multi-modality cross-media retrieval.
We observe that the methods proposed with semantic information such as CCA+SMN, HSNN, LGCFL, ml-CCA, mv-CCA and JGRHML achieve better results than CCA, CFA and BITR, which only consider the pairwise correlations. DCMIT performs better than CCA due to the use of DNN. CMCP and JRL achieve better results, for the reason that CMCP considers not only the positive but also the negative correlations among different media types, and JRL incorporates the sparse and semi-supervised regularizations to enrich the training set as well as make the solution smooth. S{\textsuperscript 2}UPG achieves the best results because it adopts the media patches to model fine-grained correlations, and the unified hypergraph can jointly model data from all media types, so as to fully exploit the correlations among them.

Table \ref{table:crossretireval} shows the MAP scores of bi-modality cross-media retrieval. Generally speaking, CMCP, HSNN, JGRHML, JRL and S{\textsuperscript 2}UPG get much better results than other methods such as BITR, CCA and CCA+SMN. 
The trends among them are different from the results on multi-modality cross-media retrieval. For example, the results of CMCP, JGRHML and JRL are close to each other on bi-modality cross-media retrieval, while JRL clearly outperforms CMCP and JGRHML on multi-modality cross-media retrieval. S{\textsuperscript 2}UPG still achieves the best results, because the fine-grained information of different media types can be modeled into one unified hypergraph to make them boost each other. It is noted that because Clickture dataset provides no category labels for supervised training, we perform unsupervised methods to verify their effectiveness, and the results are shown in Table \ref{tab:Clickture}. The overall trends among different methods on Clickture dataset are similar with other datasets.
\begin{table}[tbp]
  \centering
  \caption{The DCG@25 scores on Clickture dataset.}
    \begin{tabular}{c|c|c|c|c}
    \hline
     Dataset & BITR  & CCA  & CFA   & DCMIT \\
    \hline\hline
     Clickture & 0.474  & 0.484   & 0.486  & 0.492 \\
    \hline
    \end{tabular}%
  \label{tab:Clickture}%
   \vspace{-3mm}
\end{table}%

We also conduct experiments on Wikipedia, XMedia and Clickture datasets with the BoW features for texts, and the CNN features for images and videos, to show the performance with different features. We use the 4,096-dimensional CNN features extracted by the fc7 layer of AlexNet, and the 3,000-dimensional BoW text features. Due to page limitation, we just present the average of all MAP scores for cross-modality and bi-modality cross-media retrieval tasks on Wikipedia dataset in Table \ref{tab:wikifeatures}, and the detailed results along with results on other datasets can be found on our website: \href{http://www.icst.pku.edu.cn/mipl/XMedia}{http://www.icst.pku.edu.cn/mipl/XMedia}.  Table \ref{tab:wikifeatures} shows that features have significant impacts on the retrieval accuracy. Generally speaking, CNN features significantly improve the performance of most compared methods, while the performance of BoW features is not stable.

\begin{table*}[htbp]
  \centering
  %\tiny
  \caption{Average of all MAP scores for multi-modality and bi-modality cross-media retrieval with different features on Wikipedia dataset.}
    \begin{tabular}{c|c|c|c|c|c|c|c|c|c|c|c|c|c}
    \hline
    Image & Text & BITR  & CCA   & CCA+SMN & CFA   & CMCP  & HSNN  & JGRHML & JRL   & LGCFL & ml-CCA & mv-CCA & {S{\textsuperscript 2}UPG} \\
    \hline\hline
    BoVW  & LDA   & 0.205  & 0.254  & 0.282  & 0.257  & 0.324  & 0.313  & 0.329  & 0.335  & 0.288  & 0.277  & 0.281  & 0.360  \\
    BoVW  & BoW   & 0.118  & 0.125  & 0.129  & 0.222  & 0.345  & 0.342  & 0.258  & 0.357  & 0.320  & 0.301  & 0.193  & 0.357  \\
    CNN   & LDA   & 0.272  & 0.193  & 0.197  & 0.378  & 0.447  & 0.431  & 0.428  & 0.452  & 0.420  & 0.363  & 0.210  & 0.459  \\
    \hline
    \end{tabular}%
  \label{tab:wikifeatures}%
     \vspace{-2mm}
\end{table*}%

\section{Challenges and Open Issues}

%\begin{table*}[htbp]
%  \centering
%  \caption{Running time of cross-media retrieval task for the compared methods on Wikipedia dataset.}
%    \begin{tabular}{c|c|c|c|c|c|c|c|c|c|c|c|c|c}
%    \hline
%    Methods & BITR  & CCA   & CCA+SMN   & CFA   & CMCP  & DCMIT  & HSNN  & JGRHML   & JRL   & LGCFL & ml-CCA & mv-CCA & {S{\textsuperscript 2}UPG} \\
%    \hline
%    \hline
%    Time  & 94.1s & 0.8s & 1.6s & 0.9s & 19.4s & 0.8s & 7.5s & 9.7s & 10.1s & 0.7s & 296.6s & 0.8s & 24.1s \\
%    \hline
%    \end{tabular}%
%  \label{tab:time}%
%\end{table*}%
\label{section:challenges}

\emph{ \textbf{Dataset Construction and Benchmark Standardization}}. Datasets are very important for experimental evaluation, but as discussed in Section \ref{section:dataset}, nowadays there are only a few datasets publicly available for cross-media retrieval. Existing datasets still have shortcomings on the size, the number of media types, the rationality of categories, etc. For example, the sizes of Wikipedia and XMedia datasets are small, and Wikipedia dataset consists of only two media types (image and text). To construct high-quality datasets, specific problems should be considered such as: What categories should be included in the datasets? How many media types should be involved? How large should the dataset be? These questions are important for evaluation on the datasets. For instance, as discussed in Section \ref{section:dataset}, the high-level semantic categories of Wikipedia dataset may lead to semantic overlaps and confusions, 
which limits the objectivity of evaluation.
%so the effectiveness of methods cannot be fully and fairly evaluated.

To address the above problems, we are constructing a new dataset named \emph{XMediaNet}, which consists of five media types (text, image, video, audio and 3D model). We select 200 categories from WordNet \cite{DBLP:journals/cacm/Miller95} for ensuring the category hierarchy. These categories consist of two main parts: animals and artifacts. There are 48 kinds of animals such as elephant, owl, bee and frog as well as 152 kinds of artifacts such as violin, airplane, shotgun and camera. The total number of media instances will exceed 100,000, and they are crawled from famous websites as Wikipedia, Flickr, YouTube, Findsounds, Freesound and Yobi3D.
% and here is some information on media instances in this new dataset: 
%\begin{itemize}
%\item Text: Text paragraphs extracted from several Wikipedia articles whose topics belong to the category.
%\item Image: Pictures including the objects of the category from Flickr.
%\item Video: Video clips from YouTube with the objects of the category, whose average duration is about 100 seconds.
%\item Audio: Audio clips containing sounds made by the objects of the category from Findsounds and Freesound, such as %dog barking, clock alarm, keyboard typing and so on.
%\item 3D Model: 3D models representing the objects belonging to the category from Yobi3D.
%\end{itemize}
%In order to make the dataset more closely fit the practical scenarios, we plan to include some context relations in it.
Once the dataset is ready, we will release it on our website: \href{http://www.icst.pku.edu.cn/mipl/XMedia}{http://www.icst.pku.edu.cn/mipl/XMedia}.
We will also provide the experimental results on widely-used datasets, and encourage researchers to submit their results for building up a continuously updated benchmark (as the website of LFW face dataset \cite{LFWTech} at \href{http://vis-www.cs.umass.edu/lfw}{http://vis-www.cs.umass.edu/lfw}, and the website of ImageNet dataset \cite{DBLP:conf/cvpr/DengDSLL009} at \href{http://www.image-net.org}{http://www.image-net.org}). Researchers can directly adopt the experimental results to evaluate their own methods, which will help them focus on algorithm design, rather than the time-consuming compared methods and results, thus greatly facilitate the development of cross-media retrieval.

\emph{ \textbf{Improvement of Accuracy and Efficiency}}.
The effective yet efficient methods are still required for cross-media retrieval. First, the accuracy needs to be improved. On the one hand, existing methods still have potential to be improved. For example, graph-based methods of cross-media similarity measurement may use more context information for the effective graph construction such as link relationships. On the other hand, the discriminative power of single-media features is also important. For instance, in the experiments of Section \ref{section:experiment}, state-of-the-art methods generally adopt the low-dimensional features (e.g., 128-dimensional BoVW histogram features for image and 10-dimensional LDA features for text as in \cite{RasiwasiaMM10SemanticCCA, ZhaiTCSVT2014JRL, PengHypergraph2015}). As discussed in Section \ref{section:Result}, when more discriminative features are adopted such as CNN features for image, the retrieval accuracy will be improved.
Second, the efficiency is also an important factor for evaluations and applications. Cross-media retrieval datasets are still small-scale and limited on the number of media types up to now. Although there have been some hashing methods for cross-media retrieval as \cite{zhen2012co, hu2014iterative, zhen2012probabilistic}, the issue of efficiency has not been paid enough attention to. In the future, with the release of our large-scale XMediaNet dataset, it will be more convenient for researchers to evaluate the efficiency of their methods, which will facilitate the development of practical applications for cross-media retrieval.

%(1) accuracy: the common space learning and similarity measure. High-dimensional features, and deep networks. (2) efficiency. The index structure to support the retrieval of big data on cross-media data.

\emph{ \textbf{Applications of Deep Neural Network}}. %DNN is a generic model for information retrieval, and similar architectures have been adopted for modeling different media types. 
DNN is designed to simulate the neuronal structure of human brain which can naturally deal with the correlations of different media types, so it is worth a try to exploit DNN for bridging the ``media gap". Actually, there have been some attempts (such as the aforementioned methods \cite{zhang2014start, srivastava42012multimodal} in Section \ref{section:DnnMethods}), but they are relatively straight-forward applications of DNN, which mostly take the single-media features as raw inputs, and perform common space learning for them by extending existing models such as autoencoders. Although DNN-based methods have achieved considerable progress on cross-media retrieval \cite{DBLP:conf/ijcai/PengHQ16}, there is still potential for further improvement. The applications of DNN remain research hotspots on cross-media retrieval, as is the case with single-media retrieval.
%On the one hand, they largely depend on the input of single-media features, and are mostly limited to two media types.
On the one hand, existing methods mainly take the single-media features as inputs, so they heavily depend on the effectiveness of features. Research efforts may be devoted to designing end-to-end architectures for cross-media retrieval, which take the original media instances as inputs (e.g., the original images and audio clips), and directly get the retrieval results with DNN. Some special networks for specific media types (e.g., R-CNN for object region detection \cite{JiangDeep2015}) could also be incorporated into the unified framework of cross-media retrieval.
On the other hand, most of the existing methods are designed for only two media types. In the future works, researchers could focus on jointly analyzing more than two media types, which will make the applications of DNN in cross-media retrieval more flexible and effective.

\emph{ \textbf{Exploitation of Context Correlation Information}}.
The main challenge of cross-media retrieval is still the heterogeneous forms of different media types. Existing methods attempt to bridge the ``media gap", but only achieve limited improvement and the retrieval results are not accurate when dealing with the real-world cross-media data. The cross-media correlations are often with the context information. 
For example, if an image and an audio clip are from two web pages with link relationship, they are likely to be relevant to each other. Many existing methods (e.g., CCA, CFA and JRL) only take the co-existence relationships and semantic category labels as training information, but ignore rich context information. Actually, cross-media data on the Internet usually does not exist separately, and has important context information such as link relationships. Such context information is relatively accurate, and provides important hints to improve the accuracy of cross-media retrieval. The web data is also usually divergent, so it is important to exploit the context information for complex practical applications. We believe that researchers will pay more attention to rich context information to boost the performance of cross-media retrieval in the future works.

\emph{ \textbf{Practical Applications of Cross-media Retrieval}}.
With the constant improvement on both effectiveness and efficiency, practical applications of cross-media retrieval will become possible. These applications can provide more flexible and convenient ways to retrieve from the large-scale cross-media data, and users will like to adopt the cross-media search engine which is capable of retrieving various media types as text, image, video, audio, and 3D model with one query of any media type. In addition, other possible application scenarios include the enterprises involving the cross-media data, such as TV stations, media corporations, digital libraries and publishing companies. Both Internet and relevant enterprises will have the huge requirements of cross-media retrieval.

%\emph{ \textbf{The Research of Cross-media Reasoning}}.
%As discussed in our recent survey on cross-media analysis and reasoning \cite{DBLP:journals/jzusc/PengZZXHLZHG17}, cross-media reasoning is a new and inspiring direction for cross-media intelligence. Traditional reasoning methods are mainly text-based, which perform reasoning under fully defined premises, while knowledge and reasoning process in the real world usually involves collaboration among language, vision, and other media types. The aim of cross-media reasoning is to extend traditional text-based reasoning methods to cross-media scenarios. To achieve this, the future research should focus on directions such as cross-media knowledge graph construction and learning, and cross-media knowledge evolution and reasoning. There are limited studies on cross-media reasoning so far, but it is an important future direction, which is the key to develop practical cross-media intelligence, and will greatly benefit various applications such as cross-media retrieval.

\section{Conclusion}
\label{section:conclusion}

Cross-media retrieval is an important research topic which aims to deal with the ``media gap" for performing retrieval across different media types. 
This paper has reviewed more than 100 references to present an overview of cross-media retrieval, for building up the evaluation benchmarks, as well as facilitating the relevant research.  
Existing methods have been introduced mainly including the common space learning and cross-media similarity measurement methods. Common space learning methods explicitly learn a common space for different media types to perform retrieval, while cross-media similarity measurement methods directly measure cross-media similarities without a common space. 
The widely-used cross-media retrieval datasets have also been introduced, including  Wikipedia, XMedia, NUS-WIDE, Pascal VOC 2007 and Clickture Datasets. Among these datasets, XMedia which we have constructed is the first dataset with five media types for comprehensive and fair evaluation. We are further constructing a new dataset XMediaNet with five media types and more than 100,000 instances. 
The cross-media benchmarks, such as datasets, compared methods, evaluation metrics and experimental results have been given, and we have established a continuously updated website to present them.
Based on the discussed aspects, the main challenges and open issues have also been presented in the future works. We hope these could attract more researchers to focus on cross-media retrieval, and promote the relevant research and applications.

\bibliographystyle{IEEEtran}


\bibliography{cite}

 \vspace{-5mm}
\begin{IEEEbiography}[{\includegraphics[width=1in,height=1.25in,clip,keepaspectratio]{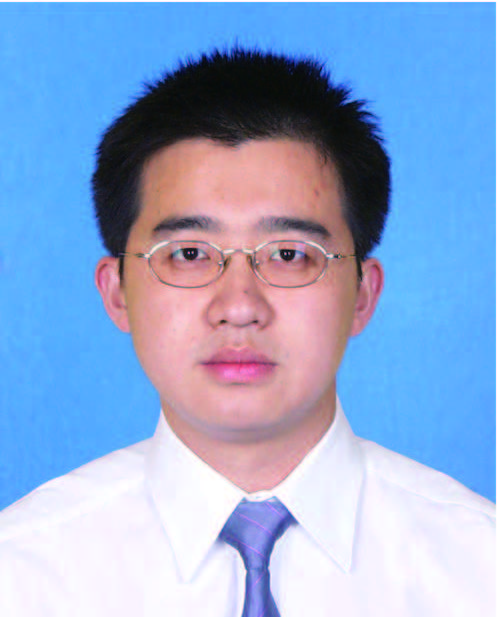}}] {Yuxin Peng}
is the professor of Institute of Computer Science and Technology (ICST), Peking University, and the chief scientist of 863 Program (National Hi-Tech Research and Development Program of China). He received the Ph.D. degree in computer application technology from Peking University in Jul. 2003. After that, he worked as an assistant professor in ICST, Peking University. He was promoted to associate professor and professor in Peking University in Aug. 2005 and Aug. 2010 respectively. In 2006, he was authorized by the ``Program for New Star in Science and Technology of Beijing" and the``Program for New Century Excellent Talents in University (NCET)". He has published over 100 papers in refereed international journals and conference proceedings, including IJCV, TIP, TCSVT, TMM, PR, ACM MM, ICCV, CVPR, IJCAI, AAAI, etc. He led his team to participate in TRECVID (TREC Video Retrieval Evaluation) many times. In TRECVID 2009, his team won four first places on 4 sub-tasks of the High-Level Feature Extraction (HLFE) task and Search task. In TRECVID 2012, his team gained four first places on all 4 sub-tasks of the Instance Search (INS) task and Known-Item Search (KIS) task. In TRECVID 2014, his team gained the first place in the Interactive Instance Search task. His team also gained both two first places in the INS task of TRECVID 2015 and 2016. Besides, he won the first prize of Beijing Science and Technology Award for Technological Invention in 2016 (ranking first). He has applied 34 patents, and obtained 15 of them. His current research interests mainly include cross-media analysis and reasoning, image and video analysis and retrieval, and computer vision.

 \vspace{-5mm}
\end{IEEEbiography}
\begin{IEEEbiography}[{\includegraphics[width=1in,height=1.25in,clip,keepaspectratio]{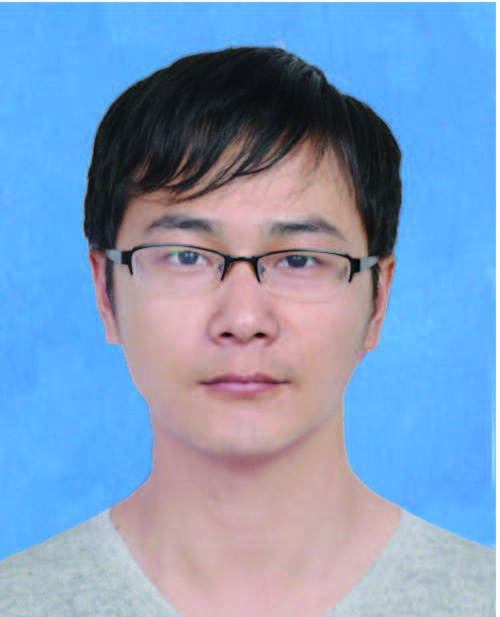}}]{Xin Huang}
 received the B.S. degree in computer science and technology from Peking University in Jul. 2014. He is currently pursuing the Ph.D. degree in the Institute of Computer Science and Technology (ICST), Peking University. His research interests include cross-media analysis and reasoning, and machine learning.
  %\vspace{-5mm}
\end{IEEEbiography}
\begin{IEEEbiography}[{\includegraphics[width=1in,height=1.25in,clip,keepaspectratio]{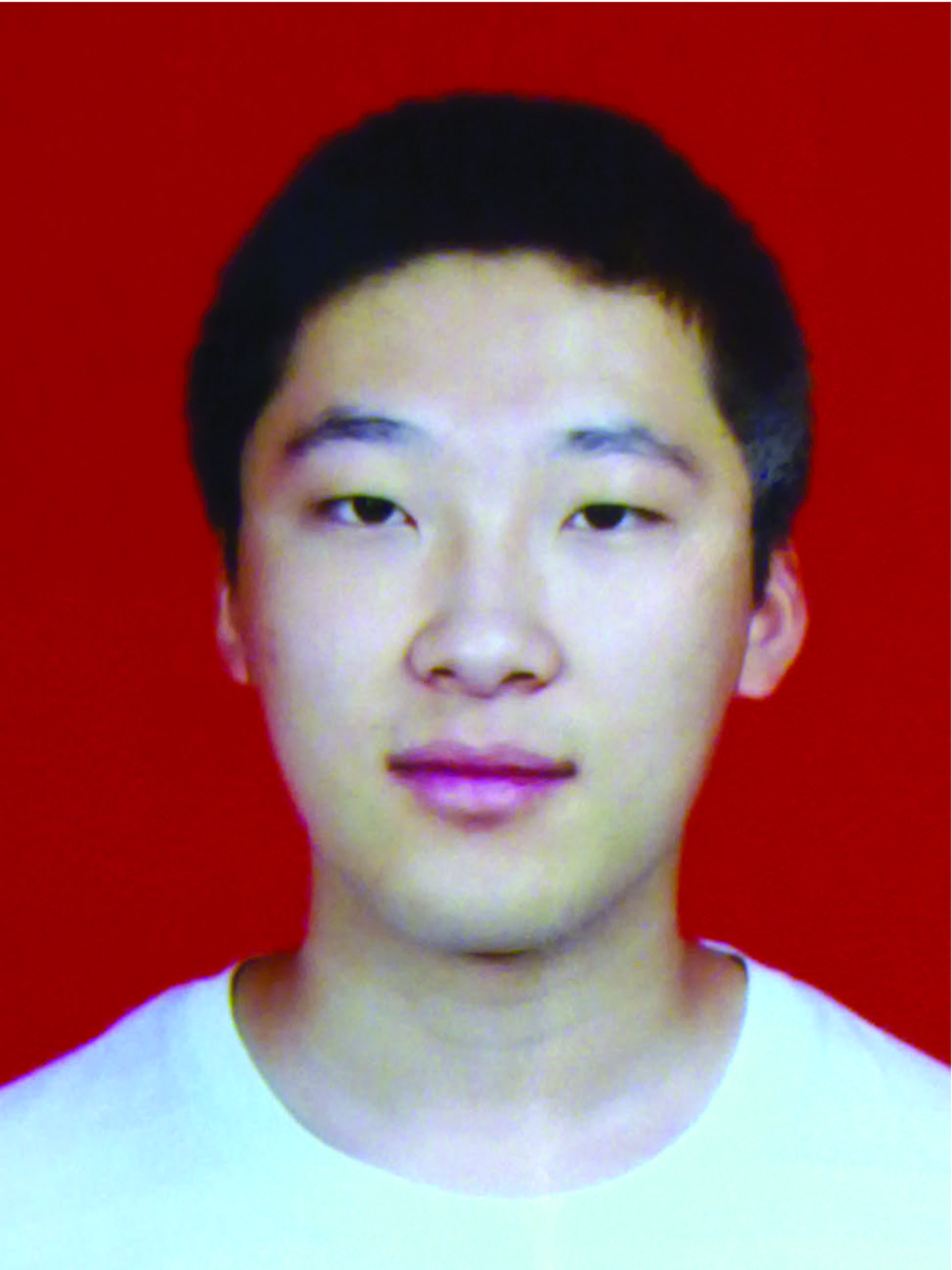}}]{Yunzhen Zhao}
received the B.S. degree in mathematical science from Peking University in Jul. 2014. He is currently pursuing the M.S. degree in the Institute of Computer Science and Technology (ICST), Peking University. His current research interests include cross-media retrieval and machine learning.
 %\vspace{-8mm}
\end{IEEEbiography}

\vfill

\end{document}